\documentclass[
 amsmath,amssymb,
 reprint,%
]{revtex4-1}

\usepackage{graphicx}
\usepackage{dcolumn}
\usepackage{bm}

\usepackage[utf8]{inputenc}
\usepackage[T1]{fontenc}
\usepackage{mathptmx}
\usepackage{etoolbox}

\makeatletter
\def\@email#1#2{%
 \endgroup
 \patchcmd{\titleblock@produce}
  {\frontmatter@RRAPformat}
  {\frontmatter@RRAPformat{\produce@RRAP{*#1\href{mailto:#2}{#2}}}\frontmatter@RRAPformat}
  {}{}
}%
\makeatother
\begin{document}

\preprint{AIP/123-QED}

\title{Non-equillibrium ultrafast optical excitation as a stimulus for ultra-small field-free magnetic skyrmions in ferrimagnetic GdFeCo}
\author{Syam Prasad P}
\author{Jyoti Ranjan Mohanty}
\email{jmohanty@phy.iith.ac.in}
\affiliation{Nanomagnetism and Microscopy Laboratory, Department of Physics, Indian Institute of Technology Hyderabad, Kandi, Sangareddy, 502284, Telangana, India}

\date{\today}

\begin{abstract}
Generating and manipulating magnetic skyrmions at ultrafast time scales is essential for future skyrmion-based racetrack memory and logic gate applications. Using the atomistic spin dynamics simulations, we demonstrate the nucleation of ultra-small field-free magnetic skyrmions in amorphous GdFeCo at picosecond timescales by femtosecond laser heating. The ultrafast nature of laser heating and subsequent cooling from a high-temperature state is crucial for forming magnetic skyrmion. The \textit{magnon localization} and \textit{magnon coalescence} are the key driving mechanisms responsible for stabilizing the magnetic skyrmions at zero field conditions. The polarization and, hence, the topological charge can be switched by exploiting the all-optical switching observed in GdFeCo. The skyrmion sizes and numbers can be controlled by varying pulse width and fluence of incident laser pulses. Applying an external magnetic field provides an additional degree of freedom to tune the skyrmion radius during the ultrafast optical creation of magnetic skyrmions. Our results provide a detailed understanding of the ultrafast creation of magnetic skyrmions using femtosecond laser pulses, a vital step in advancing next-generation skyrmion-based memory technologies.
\end{abstract}

\maketitle

\section{Introduction}
Magnetic skyrmions are nanoscale chiral spin textures having a non-trivial topology. They are stabilized in the presence of an antisymmetric exchange interaction called Dzyaloshinskii-Moriya interaction (DMI)~\cite{dzyaloshinskii1957thermodynamic,moriya1960anisotropic}, which arises in non-centrosymmetric crystals such as B20 alloys or in thin film systems where inversion symmetry is broken at the interface. Non-centrosymmetric crystals such as MnSi~\cite{muhlbauer2009skyrmion} and FeGe~\cite{yu2011near,twitchett2022confinement} exhibit the intrinsic DMI and were found to host the Bloch skyrmions. In contrast, multilayer stacks like Pt/Co/Os/Pt~\cite{tolley2018room}, Ir/Fe/Co/Pt~\cite{woo2016observation} and Pt/Co/Ta~\cite{soumyanarayanan2017tunable} host N$\acute{\text{e}}$el skyrmions. These skyrmions are stabilized by interfacial DMI, which results from broken inversion symmetry due to an interfacial layer with strong spin-orbit coupling.  
Due to their small size, topological stability, and ability to be manipulated with spin-polarized currents at low densities\cite{fert2013skyrmions,iwasaki2013current,tang2021magnetic}, magnetic skyrmions are highly promising for applications such as racetrack memory~\cite{parkin2008magnetic,fert2013skyrmions,tomasello2014strategy,wang2019controlled,perumal2023tunable}, neuromorphic computing~\cite{huang2017magnetic}, and logic gates~\cite{zhang2015magnetic,xing2016skyrmion,paikaray2022reconfigurable,tey2022chiral}. Even though magnetic skyrmions generate a lot of interest for future spintronic applications, significant challenges must be overcome to materialize skyrmion-based racetrack memory and logic gates. These challenges include room temperature stability~\cite{siemens2016minimal,buttner2018theory}, larger size, and the presence of the skyrmion Hall effect~\cite{jiang2017direct,litzius2017skyrmion}. Owing to the intrinsic perpendicular magnetic anisotropy (PMA)~\cite{leamy1979microstructure,harris1992structural,P2023171158,hansen1989magnetic}, the ability to sustain DMI at higher thicknesses and to host ultra-small skyrmions at zero field~\cite{brandao2019evolution}, and potential to avoid the skyrmion Hall effect~\cite{woo2018current,hirata2019vanishing}, ferrimagnetic rare-earth transition metal (RE-TM) alloys attract significant interest for skyrmion based devices.  

The generation, annihilation, and manipulation of magnetic skyrmions have been part of intensive research from both fundamental aspects and application points of view. Several techniques have been employed to generate and manipulate skyrmions, viz., external magnetic field~\cite{iwasaki2013universal,zhang2018manipulation}, spin-polarized current~\cite{tchoe2012skyrmion,romming2013writing}, electric field control~\cite{lemesh2018current,hsu2017electric}, voltage-controlled magnetic anisotropy~\cite{xia2018skyrmion}, DMI-gradient~\cite{gorshkov2022dmi}, etc. There is still a need to develop faster and more energy-efficient methods for creating and manipulating skyrmions. Recently, ultrafast control of magnetization using femtosecond laser pulses has gained great interest due to its ultrafast and localized nature~\cite{beaurepaire1996ultrafast}. Laser-induced magnetization dynamics and all-optical switching have been subject to many experimental and theoretical investigations~\cite{stanciu2007all,radu2011transient,ostler2012ultrafast,mangin2014engineered,jakobs2021unifying,gweha2024gd}. Advances in ultrafast magnetism have opened up a new energy-efficient technique to create and manipulate magnetic skyrmions at picosecond and sub-nanosecond time scales. In particular, several experimental and theoretical investigations have demonstrated that localized ultrafast heating from femtosecond laser pulses can induce the nucleation of magnetic skyrmions. Koshibae et al. theoretically proposed the formation of magnetic skyrmions and antiskyrmions at subnanosecond timescales followed by laser irradiation~\cite{koshibae2014creation}. Also, the ultra-fast laser-induced nucleation of skyrmion bubbles was realized experimentally in ferromagnetic CoFeB~\cite{je2018creation} and ferrimagnetic TbFeCo~\cite{finazzi2013laser}. However, these studies lack the magnetization dynamics and the information on the time scales at which skyrmion nucleation occurs. Subsequently, the skyrmion lattice in FeGe was directly visualized over a microsecond timeframe using pump-probe Lorentz transmission electron microscopy (TEM)~\cite{berruto2018laser}. Nevertheless, the poor time resolution of Lorentz-TEM restricts the mapping of magnetization dynamics at picoseconds timescales. Later, in 2021, B$\ddot{\text{u}}$ttner et al. demonstrated the skyrmion nucleation at a few hundred picosecond timescales using a single laser pulse in MgO/CoFeB/Pt and Co/Pt multilayers~\cite{buttner2021observation}. Recently, Zhang et al. reported the formation of skyrmions by the laser-induced spin reorientation transition~\cite{zhang2023optical}.  But still, the microscopic evidence of intermediate states during skyrmion nucleation remains elusive.

Atomistic spin dynamics (ASD) simulation has been proven to be an efficient tool for studying ultrafast laser-induced magnetization dynamics, and it can provide an accurate description of the dynamics of spin systems at femtosecond to picosecond time scales~\cite{iacocca2019spin,barker2013two, P2023170701}. P Olleros-Rodriguez et al. demonstrated the nucleation of metastable skyrmion lattice in ferromagnetic Pt/Co(3ML)/HM trilayer followed by laser heating using ASD simulations and proved non-equilibrium ultrafast heating is responsible for the formation of skyrmion lattice~\cite{olleros2022non}. As mentioned earlier, the ferrimagnetic RE-TM alloys provide a favourable environment to host ultra-small magnetic skyrmions for several reasons, and they can avoid the skyrmion Hall effect near compensation temperature ($T_M$)~\cite{hirata2019vanishing}. In the present work, we demonstrate the efficient generation of ultrasmall field-free skyrmions in Pt/GdFeCo bilayer by ultrafast optical excitations using ASD simulations. The ferrimagnetic amorphous GdFeCo is a widely explored material for ultrafast laser-induced magnetization dynamics and all-optical switching studies. This study shows the nucleation of ultra-small field free skyrmions generated by the non-equilibrium heating produced by femtosecond laser pulses. Also, we predict the optimal energy requirements for the successful formation of the skyrmion phase. Our study aims to establish the role of input laser parameters and external magnetic field in the ultrafast optical creation of magnetic skyrmions.

\section{methods}
To study the ultrafast laser-induced generation of magnetic skyrmions, we used atomistic spin dynamics (ASD) simulations combined with the two-temperature model (2TM). The VAMPIRE software package was used to perform the ASD simulations. We modelled the ferrimagnetic amorphous GdFeCo as two sublattices which are antiferromagnetically coupled. To build such a two-sublattice model, we first constructed a FeCo lattice with specific parameters and introduced the Gd atoms to randomly chosen sites within the FeCo lattice. This model represents the disordered nature of amorphous GdFeCo. A Pt underlayer is inserted to introduce the interfacial DMI in the system. The simulated system is considered a square-shaped sample having a 50 nm size in the x and y directions with periodic boundary conditions, and the thickness of the GdFeCo layer is set at 0.6 nm. The total energy of the spin system is defined by the Heisenberg Hamiltonian, given by
\begin{equation}
    \mathcal{H}=-\sum_{i>j}{J_{ij}\textbf{S}_i\cdot\textbf{S}_j} -\sum_i{k_u S_z^2} -\sum_i{\mu_s \textbf{S}_i\cdot\textbf{H}_{app}}-\sum_{i>j}{\textbf{D}_{ij}\cdot(\textbf{S}_i\times\textbf{S}_j})
    \label{eqn1}
\end{equation}
where $\textbf{S}_i$ and $\textbf{S}_j$ represent the normalised spin vectors at the atomic sites \textit{i} and \textit{j} and $J_{ij}$ is the symmetric exchange interaction between the spins \textit{i} and \textit{j}. $k_u$ represents the onsite uniaxial anisotropy constant, with positive z direction as an easy axis. $\mu_s$ and $\textbf{H}_{app}$ are the atomic spin moment and external magnetic field, respectively. $\textbf{D}_{ij} = D(\textbf{z}\times\textbf{r}_{ij})$ is the DMI vector, where \textbf{z} and $\textbf{r}_{ij}$ are the unit vectors along z direction and direction pointing the relative distance between the spins \textit{i} and \textit{j}. The stochastic Landau–Lifshitz–Gilbert(s-LLG) equation calculates the time dynamics of the spin system. 
\begin{equation}
    \frac{d\textbf{S}_i}{dt} = -\frac{\gamma_i}{(1+\lambda_i^2)\mu_i} \textbf{S}_i \times [\textbf{H}_i + \lambda_i\textbf{S}_i \times \textbf{H}_i]
    \label{eqn2}
\end{equation}
where $\gamma_i$, $\lambda_i$, and $\mu_i$ denote the gyromagnetic ratio, phenomenological damping constant, and magnetic moment at each site \textit{i}, respectively. $\textbf{H}_i=-\frac{\partial \mathcal{H}}{\partial \textbf{S}_i} +\boldsymbol{\zeta}_i$ is the effective field at each site \textit{i} with a stochastic thermal field $\boldsymbol{\zeta}_i$. $\boldsymbol{\zeta}_i$ couples the spin system to the thermostat, and it is represented as a white noise term, uncorrelated in space and time with the properties
\begin{equation}
    <\boldsymbol{\zeta}_{i}(t)>=0,~~~~~~~
    <\boldsymbol{\zeta}_{i}(0)\boldsymbol{\zeta}_{i}(t)>=2\delta_{ij}\delta(t)~\frac{\mu_{i}\lambda_{i}k_{B}T_{e}}{\gamma_{i}} 
\end{equation}
The values used to parametrise the spin Hamiltonian (equation~\ref{eqn1}) are listed in Table~\ref{table1}. These values correspond to the composition of $Gd_{24}Fe_{66.5}Co_{9.5}$ mentioned in reference~\cite{ostler2012ultrafast}. The DMI values at the Pt/GdFeCo interface used in this work are $4.0\times10^{-22} J/link$ for FeCo sublattice and $-4.0\times10^{-22} J/link$ for Gd sublattices, respectively. The DMI cut-off range is defined as 0.355 nm from the interface. Within this range, DMI remains constant and exponentially decays beyond this limit.   
\begin{table}[h!]
\centering
\caption{\label{table1}List of Heisenberg Hamiltonian parameters of amorphous GdFeCo}
\begin{tabular}{ccc}
\hline
 Parameters  &Value &Units\\
\hline
 $J{_{Fe-Fe}}$ & $2.835\times10^{-21}$ & J/link \\
 $J{_{Gd-Gd}}$ & $1.26\times10^{-21}$ & J/link\\
 $J_{Fe-Gd}$ & $-1.09\times10^{-21}$ & J/link\\
 $\mu_{Fe}$ & $1.92$ &$\mu_B$\\
 $\mu_{Gd}$ & $7.63$ &$\mu_B$\\
 $k_u$& $8.07246\times10^{-24}$ & J/atom \\
 $\lambda_{Fe}=\lambda_{Gd}$ & 0.05 & \\
 $\gamma_{Fe}=\gamma_{Gd}$ & $1.76\times 10^{11}$ & $T^{-1}s^{-1}$ \\
\hline
\end{tabular}
\end{table}

The time evolution of electron and phonon temperatures is calculated from the two-temperature model (2TM). This model bridges the electron and phonon temperatures to the incident laser pulse. The 2TM is represented by a pair of coupled equations, given by,

\begin{gather}
    C_e\frac{dT_e}{dt}=-G_{ep}(T_e - T_{ph})-\kappa_e\Delta T_e+ P(t) \\
    C_{ph}\frac{dT_{ph}}{dt}=-G_{ep}(T_{ph} - T_e)
\end{gather}
where $T_e$ and $T_{ph}$ are the electron and phonon temperatures respectively. $C_e=\gamma_e T_e$ and $C_{ph}$ stand for the electron and phonon-specific heat capacities, and $G_{ep}$ is the electron-phonon coupling. $\kappa_e$ is the heat-sink coupling constant representing heat dissipation to the surroundings. $P(t)=P_0e^{-(t/\tau_p)^2}$ is the pump laser energy received by the electrons in the system, and the laser pulse is assumed to have a Gaussian energy profile with a pulse width $\tau_p$. In the present work, the 2TM uses the following parameters for the calculation of electron and phonon temperatures $\gamma=2.25\times10^2 Jm^{-3}K^{-2}$, $C_{ph}=3.1\times10^6 Jm^{-3}K^{-1}$, $G_{ep}=2.5\times10^{17} Js^{-1}m^{-3}K^{-1}$, and $\kappa_e=4\times10^9 s^{-1}$ respectively~\cite{ostler2012ultrafast}. 
\begin{figure*}
\begin{center}
\includegraphics[width=0.99\linewidth]{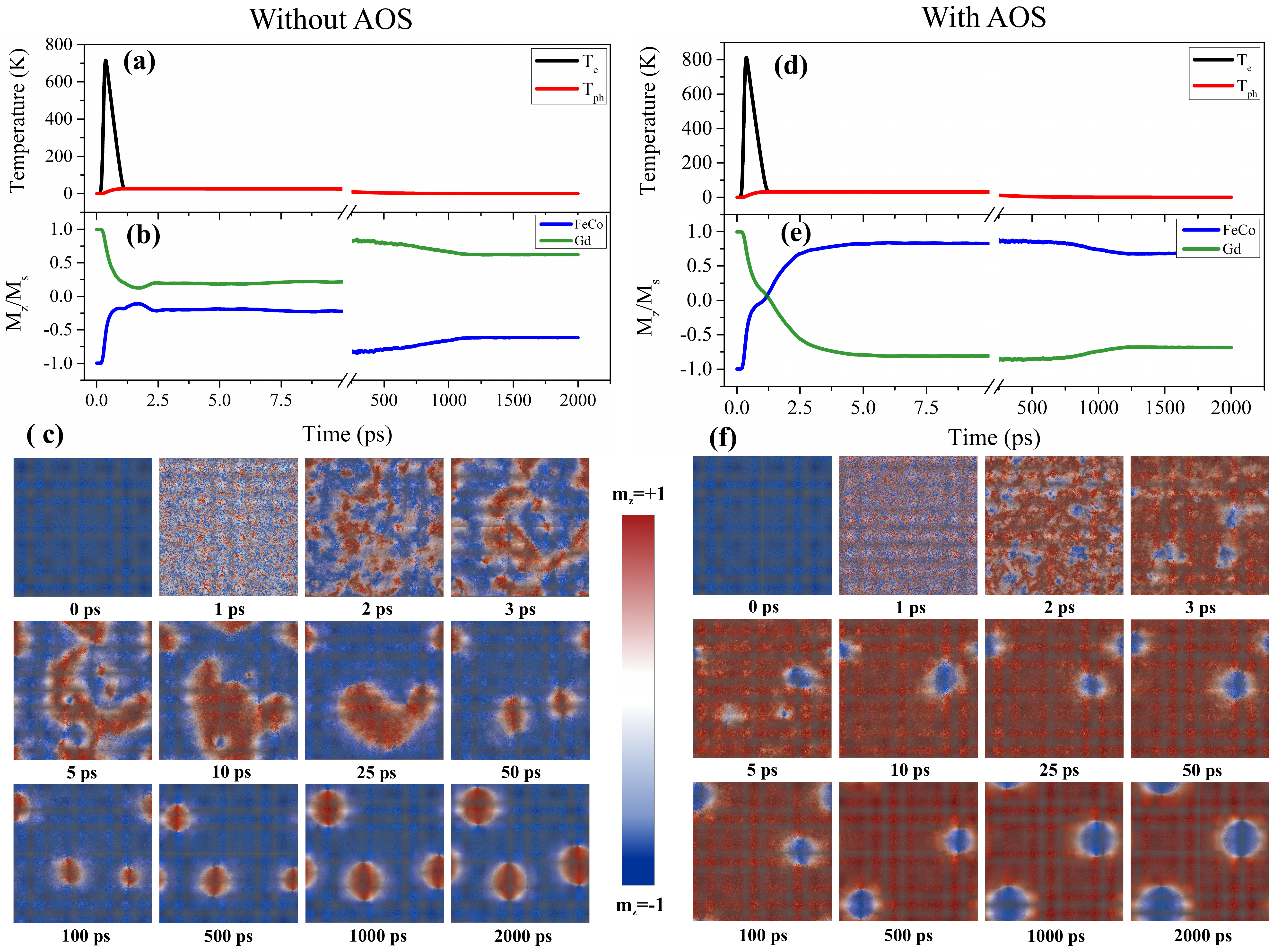}
\caption{\label{Figure_1} (a) The time dynamics of electron (black line) and phonon (red line) temperatures following the irradiation of a $100~fs$ laser pulse at a fluence of $8~mJ/cm^2$, (b) the dynamics of z-component of magnetisation of both Gd (green line) and FeCo (blue line) sublattices under the same laser parameters, indicating thermal demagnetisation, (c) the spin configurations at different intervals during the nucleation of magnetic skyrmions under the aforementioned conditions, (d) the dynamics of electron and phonon temperatures for a laser fluence of $10~mJ/cm^2$, (e)  magnetization dynamics of Gd and FeCo sublattices for the same fluence, demonstrating all-optical switching, and (f) the spin configurations at different intervals after the irradiation of $10~mJ/cm^2$ laser pulse.}
\end{center}
\end{figure*}
\section{Results} 
\subsection{Ultrafast optical creation of magnetic skyrmions at zero magnetic fields}
In the present work, using ASD simulations, we demonstrate that ultrafast heating from femtosecond laser pulses can successfully nucleate zero-field magnetic skyrmions. The nucleation protocol adopted in our simulation is as follows. We start with the initial ferrimagnetic state with Gd moments points in the positive z direction and FeCo moments in the negative z direction. The system is equilibrated at 0K for 10 ps. Next, we irradiate the system with a linearly polarised single ultrashort laser pulse having a fluence $F_0$ and pulse width $\tau_p$. When the laser pulse impinges on the system, electrons in the system quickly absorb the energy, and the laser excites a few electrons to a highly non-equilibrium state. At this stage, excited electrons have a nonthermal distribution, which Fermi-Dirac statistics cannot explain. In the following few hundred fs, a mutual thermalization occurs between the electrons, allowing them to equilibrate with each other. This leads to a very high electronic temperature where the distribution can be well understood by Fermi-Dirac statistics. Subsequently, the hot electrons transfer the thermal energy to spin and lattice and attain an equilibrium temperature. Further, the system is relaxed to 0K by dissipating energy to surroundings. The system is relaxed to 0K at a larger timescale of 2000 ps to get stable skyrmionic configurations. The rapid heating and subsequent sudden cooling from an electronic temperature above Curie temperature~($T_c$), under the action of ultrafast laser pulses, lead to the nucleation of N$\acute{\text{e}}$el type skyrmions. To characterize the topology of the laser-induced skyrmion phase, we calculate the total topological charge Q of the final spin configuration. In a continuous-field-approximation, Q is defined as~~$Q=\frac{1}{4\pi}\int\textbf{S}\cdot(\frac{\partial\textbf{S}}{\partial x}\times \frac{\partial\textbf{S}}{\partial y})dx dy$. Whereas in the case of discrete lattice, Q is determined by summing spherical areas associated with sets of three neighbouring spins arranged in a triangular configuration~\cite{rozsa2016complex}. For a skyrmion lattice having a well-defined chirality of domain walls and a unique skyrmion polarization ($P_{core}=\pm 1$), the total topological charge is characterized by the number of skyrmions $N_{sk}$ and is given by~~$Q=N_{sk}\cdot P_{core}$~\cite{olleros2022non}. The value of $P_{core}$ is taken as +1 for the skyrmion core pointing parallel to the initial magnetization direction, and it is taken as -1 for the core pointing antiparallel to the initial magnetization direction. 
\begin{figure}
\begin{center}
\includegraphics[width=0.98\linewidth]{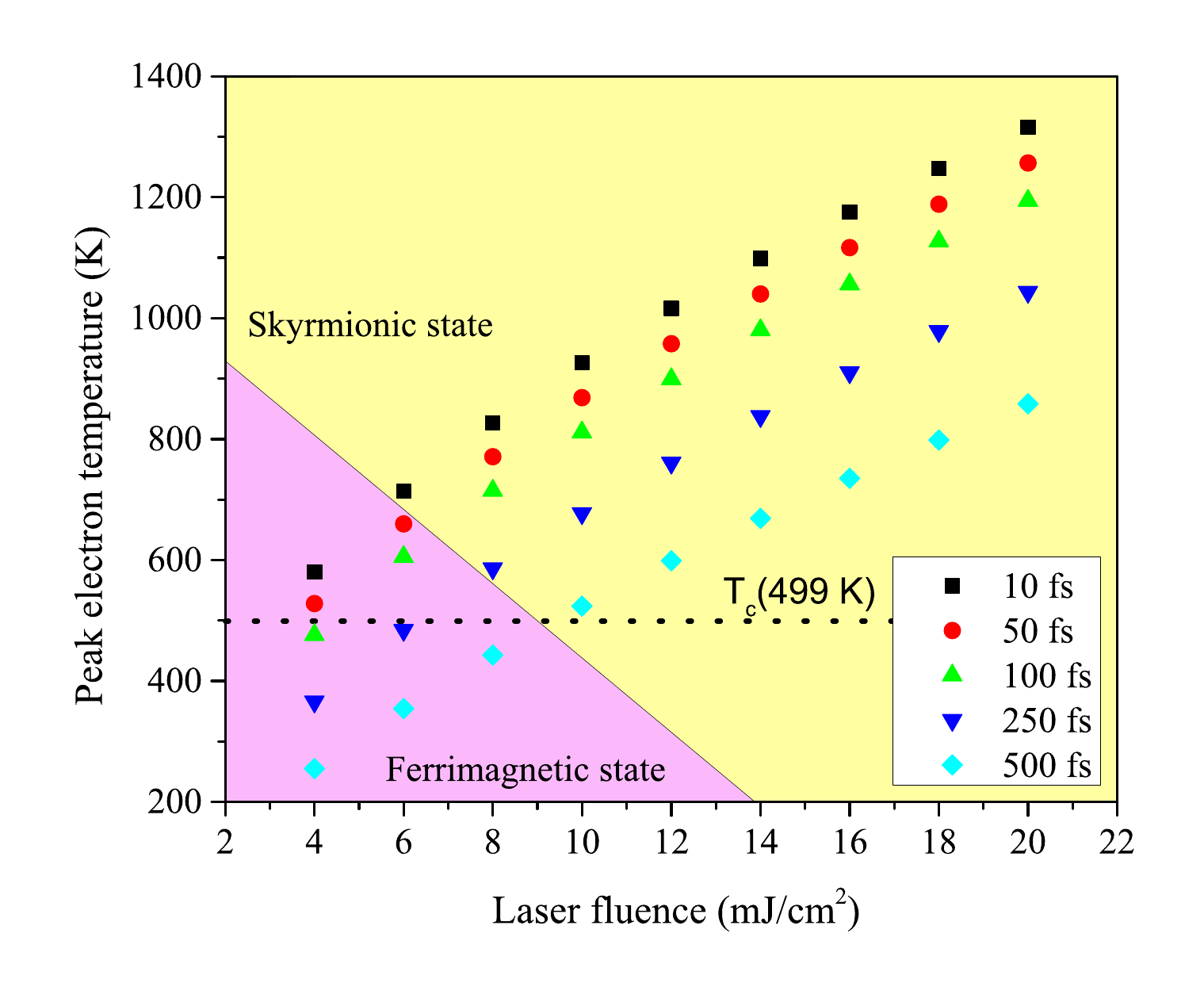}
\caption{\label{Figure_2} The peak electronic temperatures as a function of both pulse width and laser fluence. The pink region represents the fluence values with a saturated ferrimagnetic state, and the yellow region represents the fluence values with a skyrmionic state. The dotted horizontal line at 499 K represents the Curie temperature of GdFeCo.}
\end{center}
\end{figure}

The detailed mechanism of skyrmion nucleation induced by a femtosecond laser pulse is depicted in Figure~\ref{Figure_1}. Figures~\ref{Figure_1}(a) and ~\ref{Figure_1}(b) present the dynamics of electron and phonon temperatures, as well as the magnetization of Gd and FeCo sublattices, respectively, under the influence of a 100 fs laser pulse with a fluence of 8 $mJ/cm^2$. The magnetization dynamics show a thermal demagnetization at timescales of a few hundreds of femtoseconds. Figure~\ref{Figure_1}(c) illustrates the corresponding spin configurations at various intervals. Upon application of the femtosecond laser pulse, the system is excited into highly non-equilibrium electronic states. Typically, electron thermalization occurs within a few hundred femtoseconds, depending on the pulse width. In the case of a 100 fs, 8 $mJ/cm^2$ laser pulse, the electron temperature ($T_e$) peaks at 715 K within 370 fs. These hot itinerant electrons drive the spin system into a demagnetized state. Subsequently, mutual thermalization occurs among the electron, spin, and phonon subsystems over timescales of 1 to 1.5 ps. As a result, the system cools down and begins to restore ferromagnetic ordering. Even though $T_e$ is higher than $T_c$ during the ultrafast demagnetization process, the spin correlations are not completely destroyed due to the intrinsic delay in the energy transfer between electron and spin systems. Consequently, ferromagnetic ordering is reinstated at localized sites through short-range atomic exchange interactions, leading to the formation of unstable spin textures known as \textit{magnon drops} (MD). This nucleation process, termed the \textit{magnon localization} process, transpires over a time duration of 0.3 to 2 ps. In the subsequent interval of 2 to 50 ps, the nucleated magnon drops scatter, split, and merge until equilibrium spin configurations are achieved. This process is referred to as the \textit{magnon coalescence} process~\cite{iacocca2019spin,olleros2022non}.  During this magnon coalescence process, the magnon drops acquire stability and topological protection, leading to the formation of the skyrmionic phase. These results illustrate the ultrafast heating and the subsequent cooling from the paramagnetic state are responsible for the nucleation of magnetic skyrmions under the action of the ultrafast heating generated by femtosecond laser pulses. The nucleation of magnetic skyrmions by the femtosecond laser heating is mediated by the magnon localisation and magnon coalescence processes. The laser-induced skyrmion lattice discussed in the reference is a metastable state~\cite{olleros2022non}, whereas in our case, the skyrmion phase generated by laser heating is a pure ground state. To verify this, we have very slowly cooled the system from the same peak electronic temperature observed during the laser heating and relaxed over a 2 ns timescale. We compared the energies of the final spin configurations and found them to be the same in both cases. In our case, the rapid heating and sudden cooling from a paramagnetic state will not result in the formation of any metastable state, and hence, the obtained skyrmionic phases are stable configurations.         

 Figures~\ref{Figure_1} (d) and (e) show the dynamics of electron and phonon temperatures and the magnetization dynamics of Gd and FeCo sublattices, respectively, for a fluence of $10~mJ/cm^2$. Figure~\ref{Figure_1}(f) presents the corresponding spin configurations at different intervals. When the fluence changes from $8~mJ/cm^2$ to $10~mJ/cm^2$, there is a transition from thermal demagnetization to all-optical switching (AOS). This transition is evidenced by magnetization dynamics and spin configurations. A very limited number of material systems exhibit helicity-independent AOS, and ferrimagnetic GdFeCo is a widely explored material for AOS studies. As reported in the literature, we observe an AOS in GdFeCo around 1.5 ps with the formation of the transient ferromagnetic-like state~\cite{radu2011transient,ostler2012ultrafast}. The AOS switches the skyrmion polarization in the opposite direction. In the previous case ($8~mJ/cm^2$), the skyrmion core was pointing antiparallel to the initial magnetization direction, and polarization was negative, resulting in a negative sign in topological charge. At $10~mJ/cm^2$, the magnetization of both sublattices reverses, and, as a result, skyrmion polarization also switches. Here, the core of nucleated skyrmions points parallel to the initial magnetization direction. Therefore, both skyrmion polarization and topological charge are taken as positive. We observe a switching of skyrmion polarization and hence the topological charge for the fluences on and above $10~mJ/cm^2$. We have performed the simulations for a wide range of fluence values ranging from $4~mJ/cm^2$ to $20~mJ/cm^2$ with an increment of $2~mJ/cm^2$. The same is repeated for five pulse widths: 10 fs, 50 fs, 100 fs, 250 fs, and 500 fs. We have not determined the exact value of threshold fluence to observe the AOS as it is beyond the scope of this work. Threshold fluence lies in between the range $8~to~10~mJ/cm^2$ for the pulse widths 10, 50, 100 and 250 fs. For a 500 fs laser pulse, we do not observe the AOS at $10~mJ/cm^2$. Instead, we observe AOS and topological charge switching $12~mJ/cm^2$ onwards. It is known from the all-optical switching studies that threshold fluences to observe AOS increase with an increase in pulse widths~\cite{jakobs2021unifying}, which is consistent with the obtained results. 
\begin{figure}
\begin{center}
\includegraphics[width=0.98\linewidth]{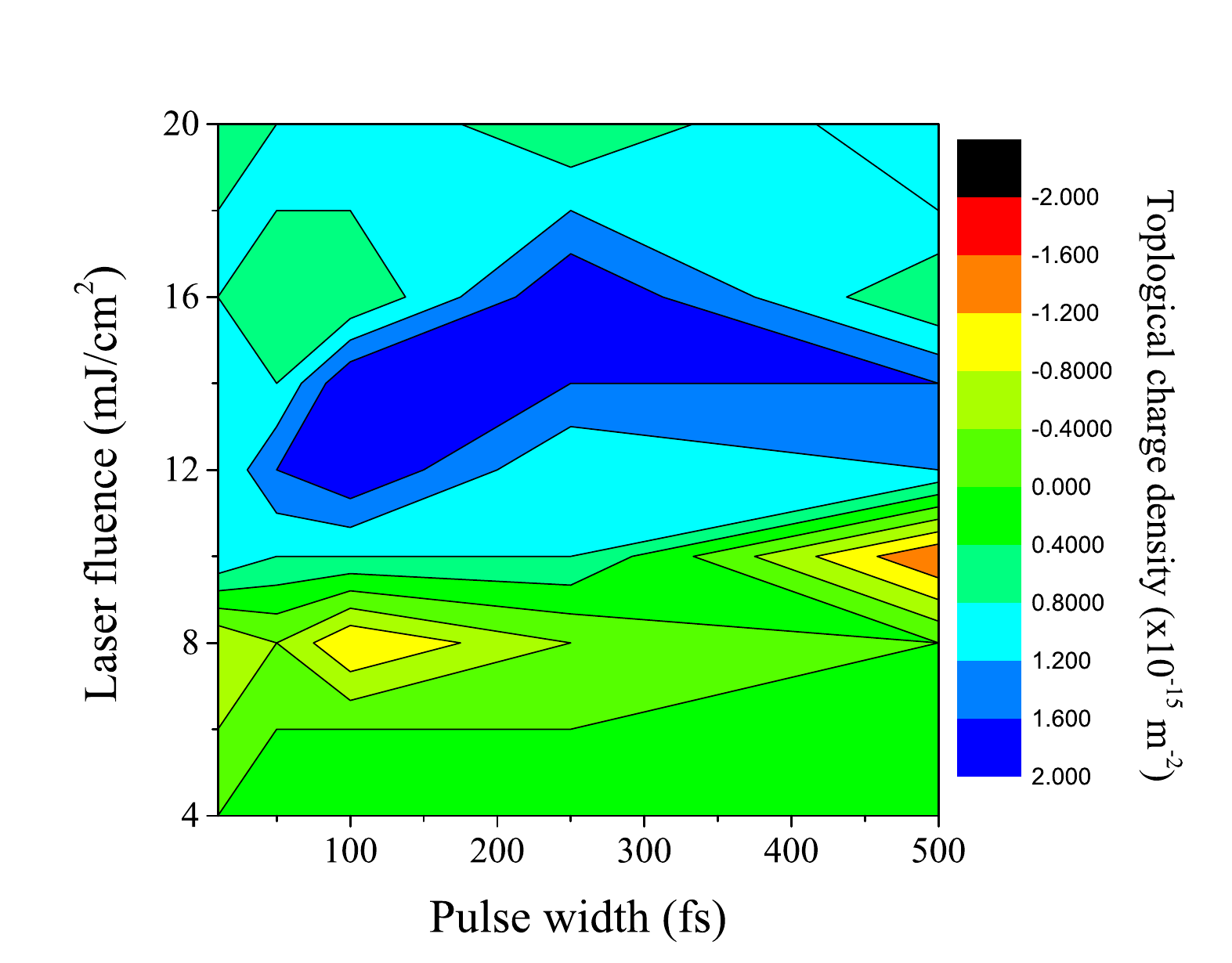}
\caption{\label{Figure_3} The topological charge density of nucleated skyrmion structures as a function of pulse width and laser fluence.}
\end{center}
\end{figure}

\begin{figure}
\begin{center}
\includegraphics[width=0.98\linewidth]{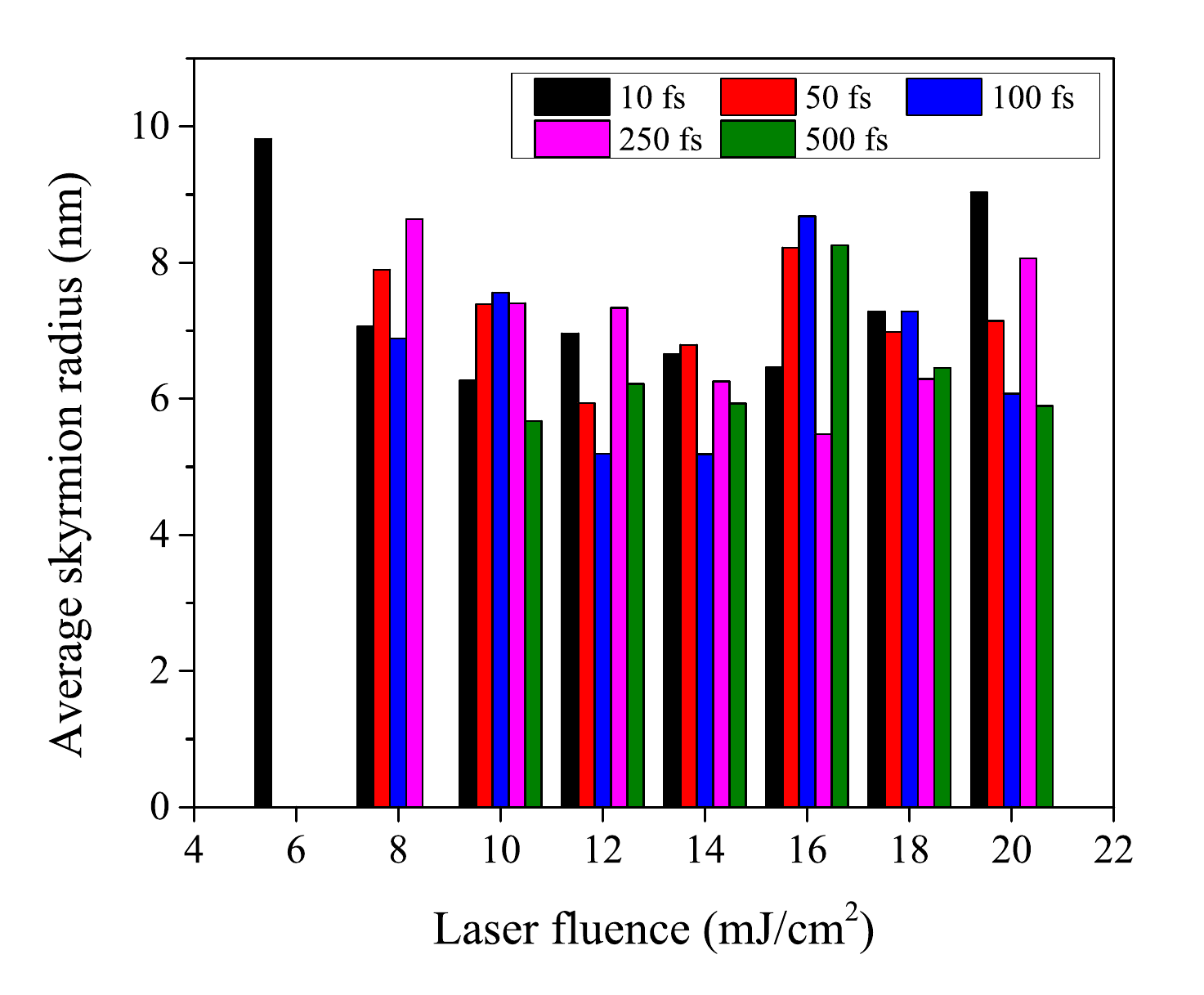}
\caption{\label{Figure_4} The average skyrmion radii of nucleated skyrmions as a function of pulse width and laser fluence.}
\end{center}
\end{figure}

 However, the ultrafast heating produced by the laser pulses is crucial for nucleating a skyrmion at zero magnetic fields; the exact role of electron temperature ($T_e$) is unclear. The peak electronic temperatures at different fluence values and pulse widths are summarised in Figure~\ref{Figure_2}. For a 10 fs laser pulse, the successful nucleation of magnetic skyrmion happens for the fluences on and above $6~mJ/cm^2$. Similarly, for the pulses with 50, 100, and 250 fs pulse widths, we observe the skyrmion nucleation for the fluences on and above $8~mJ/cm^2$. But 500 fs pulse requires a fluence of $10~mJ/cm^2$ to initiate the nucleation. The results obtained confirm that the $T_e$ needs to rapidly overcome the Curie temperature ($T_c$) for successful nucleation of magnetic skyrmions. Nevertheless, we notice that, even after exceeding the $T_c$ (499 K here), the $T_e$ is insufficient to initiate the nucleation of skyrmions at certain fluences. Successful nucleation of magnetic skyrmion requires the system to populate with a large number of magnon drops during the magnon localization process. At these fluences, the system fails to demagnetize sufficiently to produce an adequate number of magnon drops. This constraint likely arises from the inherent delay in the energy transfer between the electron and spin systems. Based on the results obtained, we infer that heating the system to a temperature above $T_c$ is a necessary but not sufficient condition to nucleate skyrmions via ultrafast optical excitations. Further, the topology of the skyrmion phase is quantified by calculating the topological charge density (q=Q/A). We constructed a phase diagram showing topological charge density q as a function of pulse width and laser fluence. Figure~\ref{Figure_3} shows the phase diagram of the topological charge density of nucleated skyrmion phases at zero field conditions. The q of the skyrmionic phase varies from $-1.6\times 10^{-15}$ to $2.0\times 10^{-15}~m^{-2}$. The maximum value of q corresponds to a $N_{sk}=~5$~in a $50\times 50~nm^2$ area. Figure~\ref{Figure_4} shows the variation of the average skyrmion radius as a function of laser fluence for different pulse widths at zero field conditions. The nucleated skyrmions are very small, and the average skyrmion radius values lie in the 5- 10 nm range under zero-field conditions. Experimental studies on the ultrafast optical generation of magnetic skyrmions suggest that skyrmion size and density are dependent on laser fluences\cite{je2018creation}. However, in our results, such a correlation is not visible. This discrepancy can be attributed to the limited dimensions of our simulated system (50 nm$\times$ 50 nm$\times$ 0.6 nm). Due to computational constraints, atomistic simulations are restricted to modelling smaller systems. Additionally, our simulations assume a uniform heating of the entire sample by the laser pulse. In contrast, in laboratory experiments, the laser spot targets a finite micron-sized area, with skyrmions expected to nucleate primarily in the central region of the spot. The relaxation dynamics in these experiments are significantly influenced by thermal gradients and magnetostatic interactions. In our simulation, the small size of the system enhances the impact of thermal fluctuations, which in turn significantly affects the relaxation dynamics and the magnon coalescence processes. This explains why we do not observe a clear trend for topological charge density and skyrmion size with laser fluence, as seen in experimental findings.  
 
\begin{figure*}[ht]
 \begin{center}
 \includegraphics[width=0.98\linewidth]{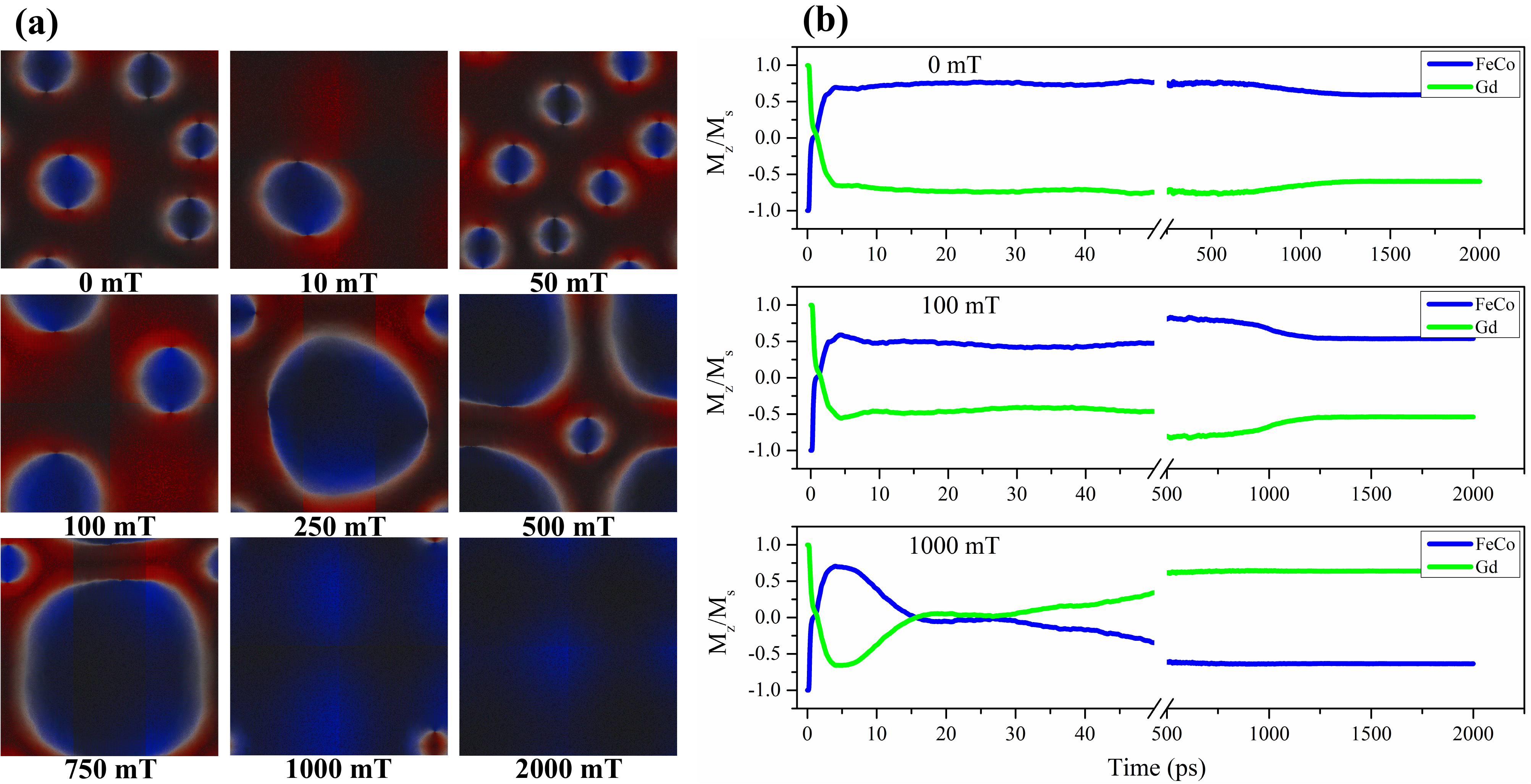}
 \caption{\label{Figure_5} (a) The final spin configurations at different magnetic fields after the application of 100 fs laser pulse having a fluence of $14~mJ/cm^2$  and (b) the dynamics of z-component of magnetization of both Gd (green line) and FeCo (blue line) at three different magnetic fields:0 mT, 100 mT and 1000 mT for the same laser parameters.}
 \end{center}
 \end{figure*}
 \subsection{Role of magnetic field on ultrafast optical creation of magnetic skyrmions}
\begin{figure}[h]
 \begin{center}
 \includegraphics[width=0.98\linewidth]{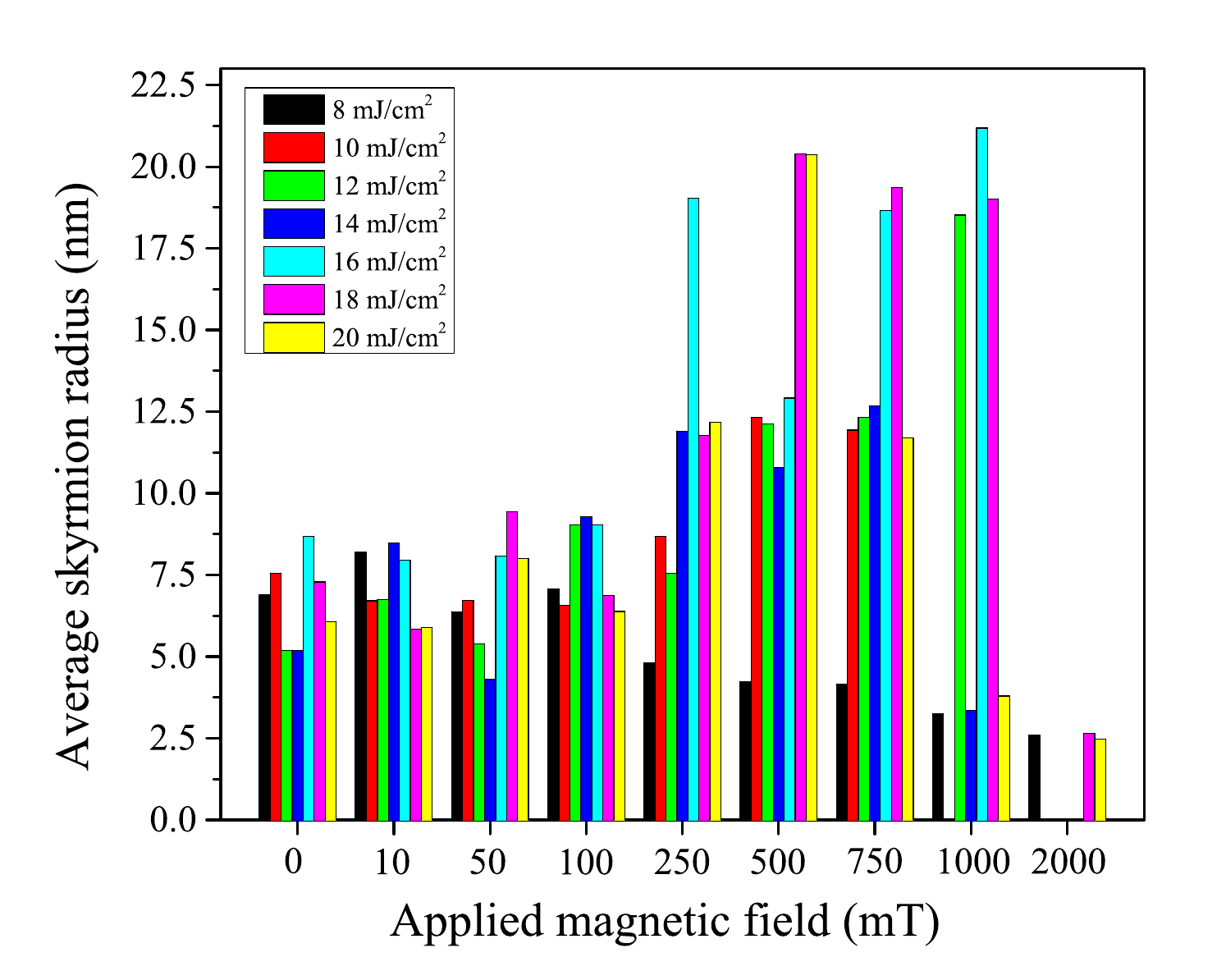}
 \caption{\label{Figure_6} The average skyrmion radii of skyrmions created using 100 fs laser pulse at different magnetic fields and fluences.}
 \end{center}
 \end{figure}

\begin{figure}[h]
 \begin{center}
 \includegraphics[width=0.98\linewidth]{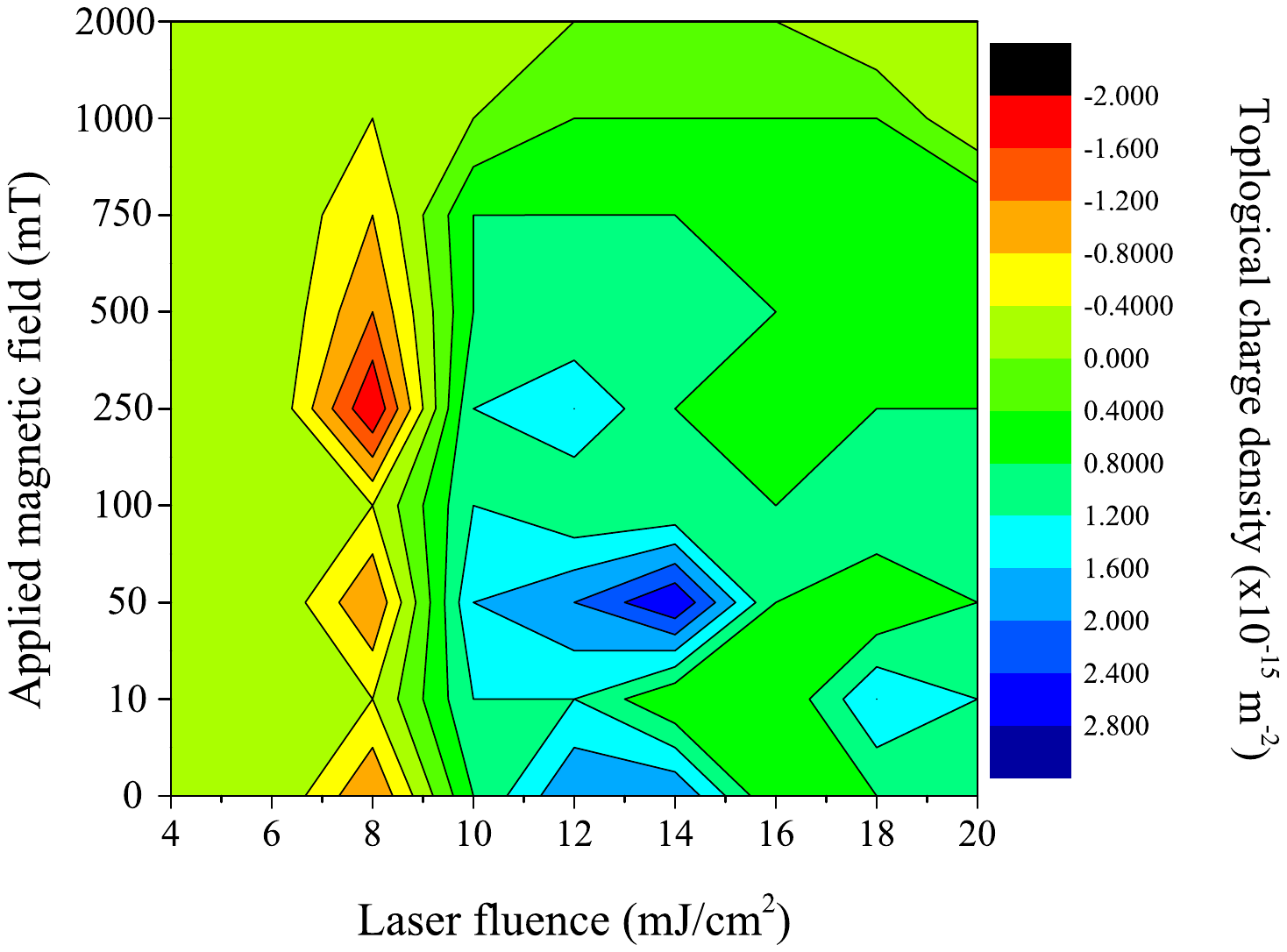}
 \caption{\label{Figure_7} The topological charge density of skyrmion structures generated using 100 fs laser pulse as a function of external magnetic field and laser fluence.}
 \end{center}
 \end{figure}

 In the previous section, we saw that the ultrafast laser pulse is a sufficient stimulus to stabilize the field-free magnetic skyrmions in ferrimagnetic GdFeCo. The nucleated skyrmions are very small, and their radii are 5-10 nm. Although the system can host skyrmions without the magnetic field followed by laser irradiation, applying an external magnetic field provides an additional degree of freedom to tune the skyrmion sizes and numbers. To understand the role of the external magnetic field during the ultrafast optical creation of magnetic skyrmions, we apply a magnetic field along the +z direction. Figure~\ref{Figure_5}(a) shows the final spin configurations in the presence of different magnetic fields after exciting the system with a 100 fs laser pulse at a fluence of $14~mJ/cm^2$. At zero field, 5 skyrmions are nucleated with an average radius of 5.2 nm. We find an increase in the average skyrmion radius when field strength increases. In this case, the average radius increases until an applied field of 750 mT. But, we find a slight deviation at 50 mT. At 50 mT field, we find 7 nucleated skyrmions with an average radius of 4.31 nm. Such deviations are found when there are more number of skyrmions. When the field is increased from 750 mT to 1 T, there is a sudden fall in skyrmion radius from 12.67 nm to 3.36 nm. No skyrmions are found at the 2T field. At this field, the strong Zeeman field forces moments to orient along the field direction, and we obtain the saturated ferrimagnetic phase. As mentioned in the previous section, the skyrmion polarization and topological charge switch if the input laser fluence exceeds the critical fluence of AOS. In this case, the fluence used is $14 mJ/cm^2$, which is sufficient to induce the AOS. Therefore, the core of nucleated skyrmion points parallel to the initial magnetization direction at lower magnetic fields. But at higher fields, the optically switched sublattice magnetization switches back to the initial magnetization direction. The switching happens when the Zeeman field is sufficiently high to overcome the anisotropy energy barrier, and switching happens around a time scale of 50 ps. As a result, the skyrmion polarization and topological charge switch at higher magnetic fields. Figure~\ref{Figure_5}(b) shows the magnetization dynamics at three different magnetic fields: 0 mT, 100 mT and 1000 mT. At 1000 mT field, we observe a field-induced re-switching of magnetization after the first laser-induced all-optical switching. Consequently, we observe a corresponding switching in skyrmion polarization from +1 to -1 at the 1000 mT field. From these results, we can conclude that both laser fluence and magnetic field can be used to control the polarization of nucleated skyrmions. The average skyrmion radius for different fields and fluences at a fixed pulse width of 100 fs is summarised in Figure~\ref{Figure_6}. The role of the magnetic field on skyrmion size is understood as follows. At zero and low magnetic fields, the size of nucleated skyrmions is small. As the magnetic field increases, the skyrmion size increases, but this increase is found up to a critical field value. This critical field varies for the different fluence values. Afterwards, the skyrmion sizes fall rapidly, and at very high fields, we obtain only a saturated ferrimagnetic state. Further, we have calculated the topology of the skyrmion phase generated in the presence of magnetic field. 
 Figure~\ref{Figure_7} shows the phase diagram representing the topological charge density as a function of magnetic field and laser fluence for a fixed pulse width of 100 fs. The topological charge density of the nucleated skyrmion phases ranges from $-2\times 10^{-15}$ to $2.8\times 10^{-15}~m^{-2}$. The results prove that applying an external magnetic field provides additional freedom to control the size and number of magnetic skyrmions during the ultrafast optical creation of magnetic skyrmions.  
 
\section{Conclusions}
In conclusion, we have developed a theoretical model based on ASD simulations to demonstrate the generation of magnetic skyrmions in amorphous GdFeCo by ultrafast laser heating. With this model, we successfully demonstrate the nucleation of magnetic skyrmions at picoseconds time scales by the ultrafast heating produced by the femtosecond laser pulses. The results show that ferrimagnetic GdFeCo can host ultra-small field-free magnetic skyrmions upon femtosecond laser heating. The ultrafast nature of heating and subsequent cooling from a high-temperature state is crucial for stabilizing the skyrmions under field-free conditions. The \textit{magnon localization} and \textit{magnon coalescence} are the key driving mechanisms responsible for stabilizing the magnetic skyrmions by femtosecond laser heating under field-free conditions. Simulations are performed for a wide range of pulse widths and fluences to estimate the optimal conditions for the successful nucleation of skyrmions, and we found that heating the system above the Curie temperature is a necessary but not sufficient condition for the formation of skyrmions. Since the ferrimagnetic GdFeCo shows the all-optical switching with a linearly polarised laser pulse, skyrmion polarisation and the topological charge can be switched by varying the laser fluence. We have calculated the sizes and topological charge densities of simulated skyrmion structures.  The skyrmion sizes and topological charge densities depend on pulse width and laser fluence. The skyrmions are small, with the average radii ranging from 5nm to 10 nm at zero field conditions. Further, we have studied the role of the magnetic field in the ultrafast optical creation of magnetic skyrmions. The skyrmion size increases as the magnetic field increases up to critical field value. Beyond this value, the size drops abruptly and at higher fields, we obtain only a saturated ferrimagnetic state. The results show that skyrmion sizes are highly tunable by applying a magnetic field. Overall, our study demonstrates the ultrafast optical creation of magnetic skyrmions using femtosecond laser pulses and efficient control of skyrmion generation by tuning the laser parameters and magnetic field. These results provide an in-depth insight into the creation and manipulation of magnetic skyrmions at ultrafast timescales using femtosecond laser pulses, which is crucial for developing next-generation skyrmion-based memories.

\begin{acknowledgments}
We acknowledge the National Supercomputing Mission (NSM) for providing computing resources of 'PARAM SEVA' at IIT, Hyderabad, which is implemented by C-DAC and supported by the Ministry of Information Technology (MeitY) and Department of Science and Technology (DST), Government of India. Syam Prasad P acknowledges IIT Hyderabad for providing the research facility and the University Grants Commission (UGC) for providing financial support.
\end{acknowledgments}
\section*{Conflict of interest}
The authors have no conflicts to disclose.
\section*{Author Contributions}
\textbf{Syam Prasad P}: Conceptualization, Methodology, Investigation, Validation, Software, Formal analysis, Visualisation, Writing – original draft. \textbf{Jyoti Ranjan Mohanty}: Discussion, Writing – review \& editing, Resources, Supervision.
\section*{Data Availability Statement}
The data that support the findings of this study are available from the corresponding author upon reasonable request.
\section*{References}
\nocite{*}
\bibliography{aipsamp}

\begin{thebibliography}{58}%
\makeatletter
\providecommand \@ifxundefined [1]{%
 \@ifx{#1\undefined}
}%
\providecommand \@ifnum [1]{%
 \ifnum #1\expandafter \@firstoftwo
 \else \expandafter \@secondoftwo
 \fi
}%
\providecommand \@ifx [1]{%
 \ifx #1\expandafter \@firstoftwo
 \else \expandafter \@secondoftwo
 \fi
}%
\providecommand \natexlab [1]{#1}%
\providecommand \enquote  [1]{``#1''}%
\providecommand \bibnamefont  [1]{#1}%
\providecommand \bibfnamefont [1]{#1}%
\providecommand \citenamefont [1]{#1}%
\providecommand \href@noop [0]{\@secondoftwo}%
\providecommand \href [0]{\begingroup \@sanitize@url \@href}%
\providecommand \@href[1]{\@@startlink{#1}\@@href}%
\providecommand \@@href[1]{\endgroup#1\@@endlink}%
\providecommand \@sanitize@url [0]{\catcode `\\12\catcode `\$12\catcode `\&12\catcode `\#12\catcode `\^12\catcode `\_12\catcode `\%12\relax}%
\providecommand \@@startlink[1]{}%
\providecommand \@@endlink[0]{}%
\providecommand \url  [0]{\begingroup\@sanitize@url \@url }%
\providecommand \@url [1]{\endgroup\@href {#1}{\urlprefix }}%
\providecommand \urlprefix  [0]{URL }%
\providecommand \Eprint [0]{\href }%
\providecommand \doibase [0]{http://dx.doi.org/}%
\providecommand \selectlanguage [0]{\@gobble}%
\providecommand \bibinfo  [0]{\@secondoftwo}%
\providecommand \bibfield  [0]{\@secondoftwo}%
\providecommand \translation [1]{[#1]}%
\providecommand \BibitemOpen [0]{}%
\providecommand \bibitemStop [0]{}%
\providecommand \bibitemNoStop [0]{.\EOS\space}%
\providecommand \EOS [0]{\spacefactor3000\relax}%
\providecommand \BibitemShut  [1]{\csname bibitem#1\endcsname}%
\let\auto@bib@innerbib\@empty
\bibitem [{\citenamefont {Dzyaloshinskii}(1957)}]{dzyaloshinskii1957thermodynamic}%
  \BibitemOpen
  \bibfield  {author} {\bibinfo {author} {\bibfnamefont {I.}~\bibnamefont {Dzyaloshinskii}},\ }\href@noop {} {\bibfield  {journal} {\bibinfo  {journal} {Sov. Phys. JETP}\ }\textbf {\bibinfo {volume} {5}},\ \bibinfo {pages} {1259} (\bibinfo {year} {1957})}\BibitemShut {NoStop}%
\bibitem [{\citenamefont {Moriya}(1960)}]{moriya1960anisotropic}%
  \BibitemOpen
  \bibfield  {author} {\bibinfo {author} {\bibfnamefont {T.}~\bibnamefont {Moriya}},\ }\href@noop {} {\bibfield  {journal} {\bibinfo  {journal} {Physical review}\ }\textbf {\bibinfo {volume} {120}},\ \bibinfo {pages} {91} (\bibinfo {year} {1960})}\BibitemShut {NoStop}%
\bibitem [{\citenamefont {M{\"u}hlbauer}\ \emph {et~al.}(2009)\citenamefont {M{\"u}hlbauer}, \citenamefont {Binz}, \citenamefont {Jonietz}, \citenamefont {Pfleiderer}, \citenamefont {Rosch}, \citenamefont {Neubauer}, \citenamefont {Georgii},\ and\ \citenamefont {B{\"o}ni}}]{muhlbauer2009skyrmion}%
  \BibitemOpen
  \bibfield  {author} {\bibinfo {author} {\bibfnamefont {S.}~\bibnamefont {M{\"u}hlbauer}}, \bibinfo {author} {\bibfnamefont {B.}~\bibnamefont {Binz}}, \bibinfo {author} {\bibfnamefont {F.}~\bibnamefont {Jonietz}}, \bibinfo {author} {\bibfnamefont {C.}~\bibnamefont {Pfleiderer}}, \bibinfo {author} {\bibfnamefont {A.}~\bibnamefont {Rosch}}, \bibinfo {author} {\bibfnamefont {A.}~\bibnamefont {Neubauer}}, \bibinfo {author} {\bibfnamefont {R.}~\bibnamefont {Georgii}}, \ and\ \bibinfo {author} {\bibfnamefont {P.}~\bibnamefont {B{\"o}ni}},\ }\href@noop {} {\bibfield  {journal} {\bibinfo  {journal} {Science}\ }\textbf {\bibinfo {volume} {323}},\ \bibinfo {pages} {915} (\bibinfo {year} {2009})}\BibitemShut {NoStop}%
\bibitem [{\citenamefont {Yu}\ \emph {et~al.}(2011)\citenamefont {Yu}, \citenamefont {Kanazawa}, \citenamefont {Onose}, \citenamefont {Kimoto}, \citenamefont {Zhang}, \citenamefont {Ishiwata}, \citenamefont {Matsui},\ and\ \citenamefont {Tokura}}]{yu2011near}%
  \BibitemOpen
  \bibfield  {author} {\bibinfo {author} {\bibfnamefont {X.}~\bibnamefont {Yu}}, \bibinfo {author} {\bibfnamefont {N.}~\bibnamefont {Kanazawa}}, \bibinfo {author} {\bibfnamefont {Y.}~\bibnamefont {Onose}}, \bibinfo {author} {\bibfnamefont {K.}~\bibnamefont {Kimoto}}, \bibinfo {author} {\bibfnamefont {W.}~\bibnamefont {Zhang}}, \bibinfo {author} {\bibfnamefont {S.}~\bibnamefont {Ishiwata}}, \bibinfo {author} {\bibfnamefont {Y.}~\bibnamefont {Matsui}}, \ and\ \bibinfo {author} {\bibfnamefont {Y.}~\bibnamefont {Tokura}},\ }\href@noop {} {\bibfield  {journal} {\bibinfo  {journal} {Nature materials}\ }\textbf {\bibinfo {volume} {10}},\ \bibinfo {pages} {106} (\bibinfo {year} {2011})}\BibitemShut {NoStop}%
\bibitem [{\citenamefont {Twitchett-Harrison}\ \emph {et~al.}(2022)\citenamefont {Twitchett-Harrison}, \citenamefont {Loudon}, \citenamefont {Pepper}, \citenamefont {Birch}, \citenamefont {Fangohr}, \citenamefont {Midgley}, \citenamefont {Balakrishnan},\ and\ \citenamefont {Hatton}}]{twitchett2022confinement}%
  \BibitemOpen
  \bibfield  {author} {\bibinfo {author} {\bibfnamefont {A.~C.}\ \bibnamefont {Twitchett-Harrison}}, \bibinfo {author} {\bibfnamefont {J.~C.}\ \bibnamefont {Loudon}}, \bibinfo {author} {\bibfnamefont {R.~A.}\ \bibnamefont {Pepper}}, \bibinfo {author} {\bibfnamefont {M.~T.}\ \bibnamefont {Birch}}, \bibinfo {author} {\bibfnamefont {H.}~\bibnamefont {Fangohr}}, \bibinfo {author} {\bibfnamefont {P.~A.}\ \bibnamefont {Midgley}}, \bibinfo {author} {\bibfnamefont {G.}~\bibnamefont {Balakrishnan}}, \ and\ \bibinfo {author} {\bibfnamefont {P.~D.}\ \bibnamefont {Hatton}},\ }\href@noop {} {\bibfield  {journal} {\bibinfo  {journal} {ACS Applied Electronic Materials}\ }\textbf {\bibinfo {volume} {4}},\ \bibinfo {pages} {4427} (\bibinfo {year} {2022})}\BibitemShut {NoStop}%
\bibitem [{\citenamefont {Tolley}\ \emph {et~al.}(2018)\citenamefont {Tolley}, \citenamefont {Montoya},\ and\ \citenamefont {Fullerton}}]{tolley2018room}%
  \BibitemOpen
  \bibfield  {author} {\bibinfo {author} {\bibfnamefont {R.}~\bibnamefont {Tolley}}, \bibinfo {author} {\bibfnamefont {S.}~\bibnamefont {Montoya}}, \ and\ \bibinfo {author} {\bibfnamefont {E.}~\bibnamefont {Fullerton}},\ }\href@noop {} {\bibfield  {journal} {\bibinfo  {journal} {Physical Review Materials}\ }\textbf {\bibinfo {volume} {2}},\ \bibinfo {pages} {044404} (\bibinfo {year} {2018})}\BibitemShut {NoStop}%
\bibitem [{\citenamefont {Woo}\ \emph {et~al.}(2016)\citenamefont {Woo}, \citenamefont {Litzius}, \citenamefont {Kr{\"u}ger}, \citenamefont {Im}, \citenamefont {Caretta}, \citenamefont {Richter}, \citenamefont {Mann}, \citenamefont {Krone}, \citenamefont {Reeve}, \citenamefont {Weigand} \emph {et~al.}}]{woo2016observation}%
  \BibitemOpen
  \bibfield  {author} {\bibinfo {author} {\bibfnamefont {S.}~\bibnamefont {Woo}}, \bibinfo {author} {\bibfnamefont {K.}~\bibnamefont {Litzius}}, \bibinfo {author} {\bibfnamefont {B.}~\bibnamefont {Kr{\"u}ger}}, \bibinfo {author} {\bibfnamefont {M.-Y.}\ \bibnamefont {Im}}, \bibinfo {author} {\bibfnamefont {L.}~\bibnamefont {Caretta}}, \bibinfo {author} {\bibfnamefont {K.}~\bibnamefont {Richter}}, \bibinfo {author} {\bibfnamefont {M.}~\bibnamefont {Mann}}, \bibinfo {author} {\bibfnamefont {A.}~\bibnamefont {Krone}}, \bibinfo {author} {\bibfnamefont {R.~M.}\ \bibnamefont {Reeve}}, \bibinfo {author} {\bibfnamefont {M.}~\bibnamefont {Weigand}},  \emph {et~al.},\ }\href@noop {} {\bibfield  {journal} {\bibinfo  {journal} {Nature materials}\ }\textbf {\bibinfo {volume} {15}},\ \bibinfo {pages} {501} (\bibinfo {year} {2016})}\BibitemShut {NoStop}%
\bibitem [{\citenamefont {Soumyanarayanan}\ \emph {et~al.}(2017)\citenamefont {Soumyanarayanan}, \citenamefont {Raju}, \citenamefont {Gonzalez~Oyarce}, \citenamefont {Tan}, \citenamefont {Im}, \citenamefont {Petrovi{\'c}}, \citenamefont {Ho}, \citenamefont {Khoo}, \citenamefont {Tran}, \citenamefont {Gan} \emph {et~al.}}]{soumyanarayanan2017tunable}%
  \BibitemOpen
  \bibfield  {author} {\bibinfo {author} {\bibfnamefont {A.}~\bibnamefont {Soumyanarayanan}}, \bibinfo {author} {\bibfnamefont {M.}~\bibnamefont {Raju}}, \bibinfo {author} {\bibfnamefont {A.}~\bibnamefont {Gonzalez~Oyarce}}, \bibinfo {author} {\bibfnamefont {A.~K.}\ \bibnamefont {Tan}}, \bibinfo {author} {\bibfnamefont {M.-Y.}\ \bibnamefont {Im}}, \bibinfo {author} {\bibfnamefont {A.~P.}\ \bibnamefont {Petrovi{\'c}}}, \bibinfo {author} {\bibfnamefont {P.}~\bibnamefont {Ho}}, \bibinfo {author} {\bibfnamefont {K.}~\bibnamefont {Khoo}}, \bibinfo {author} {\bibfnamefont {M.}~\bibnamefont {Tran}}, \bibinfo {author} {\bibfnamefont {C.}~\bibnamefont {Gan}},  \emph {et~al.},\ }\href@noop {} {\bibfield  {journal} {\bibinfo  {journal} {Nature materials}\ }\textbf {\bibinfo {volume} {16}},\ \bibinfo {pages} {898} (\bibinfo {year} {2017})}\BibitemShut {NoStop}%
\bibitem [{\citenamefont {Fert}\ \emph {et~al.}(2013)\citenamefont {Fert}, \citenamefont {Cros},\ and\ \citenamefont {Sampaio}}]{fert2013skyrmions}%
  \BibitemOpen
  \bibfield  {author} {\bibinfo {author} {\bibfnamefont {A.}~\bibnamefont {Fert}}, \bibinfo {author} {\bibfnamefont {V.}~\bibnamefont {Cros}}, \ and\ \bibinfo {author} {\bibfnamefont {J.}~\bibnamefont {Sampaio}},\ }\href@noop {} {\bibfield  {journal} {\bibinfo  {journal} {Nature nanotechnology}\ }\textbf {\bibinfo {volume} {8}},\ \bibinfo {pages} {152} (\bibinfo {year} {2013})}\BibitemShut {NoStop}%
\bibitem [{\citenamefont {Iwasaki}\ \emph {et~al.}(2013{\natexlab{a}})\citenamefont {Iwasaki}, \citenamefont {Mochizuki},\ and\ \citenamefont {Nagaosa}}]{iwasaki2013current}%
  \BibitemOpen
  \bibfield  {author} {\bibinfo {author} {\bibfnamefont {J.}~\bibnamefont {Iwasaki}}, \bibinfo {author} {\bibfnamefont {M.}~\bibnamefont {Mochizuki}}, \ and\ \bibinfo {author} {\bibfnamefont {N.}~\bibnamefont {Nagaosa}},\ }\href@noop {} {\bibfield  {journal} {\bibinfo  {journal} {Nature nanotechnology}\ }\textbf {\bibinfo {volume} {8}},\ \bibinfo {pages} {742} (\bibinfo {year} {2013}{\natexlab{a}})}\BibitemShut {NoStop}%
\bibitem [{\citenamefont {Tang}\ \emph {et~al.}(2021)\citenamefont {Tang}, \citenamefont {Wu}, \citenamefont {Wang}, \citenamefont {Kong}, \citenamefont {Lv}, \citenamefont {Wei}, \citenamefont {Zang}, \citenamefont {Tian},\ and\ \citenamefont {Du}}]{tang2021magnetic}%
  \BibitemOpen
  \bibfield  {author} {\bibinfo {author} {\bibfnamefont {J.}~\bibnamefont {Tang}}, \bibinfo {author} {\bibfnamefont {Y.}~\bibnamefont {Wu}}, \bibinfo {author} {\bibfnamefont {W.}~\bibnamefont {Wang}}, \bibinfo {author} {\bibfnamefont {L.}~\bibnamefont {Kong}}, \bibinfo {author} {\bibfnamefont {B.}~\bibnamefont {Lv}}, \bibinfo {author} {\bibfnamefont {W.}~\bibnamefont {Wei}}, \bibinfo {author} {\bibfnamefont {J.}~\bibnamefont {Zang}}, \bibinfo {author} {\bibfnamefont {M.}~\bibnamefont {Tian}}, \ and\ \bibinfo {author} {\bibfnamefont {H.}~\bibnamefont {Du}},\ }\href@noop {} {\bibfield  {journal} {\bibinfo  {journal} {Nature Nanotechnology}\ }\textbf {\bibinfo {volume} {16}},\ \bibinfo {pages} {1086} (\bibinfo {year} {2021})}\BibitemShut {NoStop}%
\bibitem [{\citenamefont {Parkin}\ \emph {et~al.}(2008)\citenamefont {Parkin}, \citenamefont {Hayashi},\ and\ \citenamefont {Thomas}}]{parkin2008magnetic}%
  \BibitemOpen
  \bibfield  {author} {\bibinfo {author} {\bibfnamefont {S.~S.}\ \bibnamefont {Parkin}}, \bibinfo {author} {\bibfnamefont {M.}~\bibnamefont {Hayashi}}, \ and\ \bibinfo {author} {\bibfnamefont {L.}~\bibnamefont {Thomas}},\ }\href@noop {} {\bibfield  {journal} {\bibinfo  {journal} {Science}\ }\textbf {\bibinfo {volume} {320}},\ \bibinfo {pages} {190} (\bibinfo {year} {2008})}\BibitemShut {NoStop}%
\bibitem [{\citenamefont {Tomasello}\ \emph {et~al.}(2014)\citenamefont {Tomasello}, \citenamefont {Martinez}, \citenamefont {Zivieri}, \citenamefont {Torres}, \citenamefont {Carpentieri},\ and\ \citenamefont {Finocchio}}]{tomasello2014strategy}%
  \BibitemOpen
  \bibfield  {author} {\bibinfo {author} {\bibfnamefont {R.}~\bibnamefont {Tomasello}}, \bibinfo {author} {\bibfnamefont {E.}~\bibnamefont {Martinez}}, \bibinfo {author} {\bibfnamefont {R.}~\bibnamefont {Zivieri}}, \bibinfo {author} {\bibfnamefont {L.}~\bibnamefont {Torres}}, \bibinfo {author} {\bibfnamefont {M.}~\bibnamefont {Carpentieri}}, \ and\ \bibinfo {author} {\bibfnamefont {G.}~\bibnamefont {Finocchio}},\ }\href@noop {} {\bibfield  {journal} {\bibinfo  {journal} {Scientific reports}\ }\textbf {\bibinfo {volume} {4}},\ \bibinfo {pages} {1} (\bibinfo {year} {2014})}\BibitemShut {NoStop}%
\bibitem [{\citenamefont {Wang}\ \emph {et~al.}(2019)\citenamefont {Wang}, \citenamefont {Qian}, \citenamefont {Ying}, \citenamefont {Xiao},\ and\ \citenamefont {Wu}}]{wang2019controlled}%
  \BibitemOpen
  \bibfield  {author} {\bibinfo {author} {\bibfnamefont {K.}~\bibnamefont {Wang}}, \bibinfo {author} {\bibfnamefont {L.}~\bibnamefont {Qian}}, \bibinfo {author} {\bibfnamefont {S.-C.}\ \bibnamefont {Ying}}, \bibinfo {author} {\bibfnamefont {G.}~\bibnamefont {Xiao}}, \ and\ \bibinfo {author} {\bibfnamefont {X.}~\bibnamefont {Wu}},\ }\href@noop {} {\bibfield  {journal} {\bibinfo  {journal} {Nanoscale}\ }\textbf {\bibinfo {volume} {11}},\ \bibinfo {pages} {6952} (\bibinfo {year} {2019})}\BibitemShut {NoStop}%
\bibitem [{\citenamefont {Perumal}\ \emph {et~al.}(2023)\citenamefont {Perumal}, \citenamefont {Sankaran~Kunnath}, \citenamefont {Priyanka},\ and\ \citenamefont {Sinha}}]{perumal2023tunable}%
  \BibitemOpen
  \bibfield  {author} {\bibinfo {author} {\bibfnamefont {H.~P.}\ \bibnamefont {Perumal}}, \bibinfo {author} {\bibfnamefont {S.}~\bibnamefont {Sankaran~Kunnath}}, \bibinfo {author} {\bibfnamefont {B.}~\bibnamefont {Priyanka}}, \ and\ \bibinfo {author} {\bibfnamefont {J.}~\bibnamefont {Sinha}},\ }\href@noop {} {\bibfield  {journal} {\bibinfo  {journal} {ACS Applied Electronic Materials}\ }\textbf {\bibinfo {volume} {5}},\ \bibinfo {pages} {3641} (\bibinfo {year} {2023})}\BibitemShut {NoStop}%
\bibitem [{\citenamefont {Huang}\ \emph {et~al.}(2017)\citenamefont {Huang}, \citenamefont {Kang}, \citenamefont {Zhang}, \citenamefont {Zhou},\ and\ \citenamefont {Zhao}}]{huang2017magnetic}%
  \BibitemOpen
  \bibfield  {author} {\bibinfo {author} {\bibfnamefont {Y.}~\bibnamefont {Huang}}, \bibinfo {author} {\bibfnamefont {W.}~\bibnamefont {Kang}}, \bibinfo {author} {\bibfnamefont {X.}~\bibnamefont {Zhang}}, \bibinfo {author} {\bibfnamefont {Y.}~\bibnamefont {Zhou}}, \ and\ \bibinfo {author} {\bibfnamefont {W.}~\bibnamefont {Zhao}},\ }\href@noop {} {\bibfield  {journal} {\bibinfo  {journal} {Nanotechnology}\ }\textbf {\bibinfo {volume} {28}},\ \bibinfo {pages} {08LT02} (\bibinfo {year} {2017})}\BibitemShut {NoStop}%
\bibitem [{\citenamefont {Zhang}\ \emph {et~al.}(2015)\citenamefont {Zhang}, \citenamefont {Ezawa},\ and\ \citenamefont {Zhou}}]{zhang2015magnetic}%
  \BibitemOpen
  \bibfield  {author} {\bibinfo {author} {\bibfnamefont {X.}~\bibnamefont {Zhang}}, \bibinfo {author} {\bibfnamefont {M.}~\bibnamefont {Ezawa}}, \ and\ \bibinfo {author} {\bibfnamefont {Y.}~\bibnamefont {Zhou}},\ }\href@noop {} {\bibfield  {journal} {\bibinfo  {journal} {Scientific reports}\ }\textbf {\bibinfo {volume} {5}},\ \bibinfo {pages} {9400} (\bibinfo {year} {2015})}\BibitemShut {NoStop}%
\bibitem [{\citenamefont {Xing}\ \emph {et~al.}(2016)\citenamefont {Xing}, \citenamefont {Pong},\ and\ \citenamefont {Zhou}}]{xing2016skyrmion}%
  \BibitemOpen
  \bibfield  {author} {\bibinfo {author} {\bibfnamefont {X.}~\bibnamefont {Xing}}, \bibinfo {author} {\bibfnamefont {P.~W.}\ \bibnamefont {Pong}}, \ and\ \bibinfo {author} {\bibfnamefont {Y.}~\bibnamefont {Zhou}},\ }\href@noop {} {\bibfield  {journal} {\bibinfo  {journal} {Physical Review B}\ }\textbf {\bibinfo {volume} {94}},\ \bibinfo {pages} {054408} (\bibinfo {year} {2016})}\BibitemShut {NoStop}%
\bibitem [{\citenamefont {Paikaray}\ \emph {et~al.}(2022)\citenamefont {Paikaray}, \citenamefont {Kuchibhotla}, \citenamefont {Haldar},\ and\ \citenamefont {Murapaka}}]{paikaray2022reconfigurable}%
  \BibitemOpen
  \bibfield  {author} {\bibinfo {author} {\bibfnamefont {B.}~\bibnamefont {Paikaray}}, \bibinfo {author} {\bibfnamefont {M.}~\bibnamefont {Kuchibhotla}}, \bibinfo {author} {\bibfnamefont {A.}~\bibnamefont {Haldar}}, \ and\ \bibinfo {author} {\bibfnamefont {C.}~\bibnamefont {Murapaka}},\ }\href@noop {} {\bibfield  {journal} {\bibinfo  {journal} {ACS Applied Electronic Materials}\ }\textbf {\bibinfo {volume} {4}},\ \bibinfo {pages} {2290} (\bibinfo {year} {2022})}\BibitemShut {NoStop}%
\bibitem [{\citenamefont {Tey}\ \emph {et~al.}(2022)\citenamefont {Tey}, \citenamefont {Chen}, \citenamefont {Soumyanarayanan},\ and\ \citenamefont {Ho}}]{tey2022chiral}%
  \BibitemOpen
  \bibfield  {author} {\bibinfo {author} {\bibfnamefont {M.~N.}\ \bibnamefont {Tey}}, \bibinfo {author} {\bibfnamefont {X.}~\bibnamefont {Chen}}, \bibinfo {author} {\bibfnamefont {A.}~\bibnamefont {Soumyanarayanan}}, \ and\ \bibinfo {author} {\bibfnamefont {P.}~\bibnamefont {Ho}},\ }\href@noop {} {\bibfield  {journal} {\bibinfo  {journal} {ACS Applied Electronic Materials}\ }\textbf {\bibinfo {volume} {4}},\ \bibinfo {pages} {5088} (\bibinfo {year} {2022})}\BibitemShut {NoStop}%
\bibitem [{\citenamefont {Siemens}\ \emph {et~al.}(2016)\citenamefont {Siemens}, \citenamefont {Zhang}, \citenamefont {Hagemeister}, \citenamefont {Vedmedenko},\ and\ \citenamefont {Wiesendanger}}]{siemens2016minimal}%
  \BibitemOpen
  \bibfield  {author} {\bibinfo {author} {\bibfnamefont {A.}~\bibnamefont {Siemens}}, \bibinfo {author} {\bibfnamefont {Y.}~\bibnamefont {Zhang}}, \bibinfo {author} {\bibfnamefont {J.}~\bibnamefont {Hagemeister}}, \bibinfo {author} {\bibfnamefont {E.}~\bibnamefont {Vedmedenko}}, \ and\ \bibinfo {author} {\bibfnamefont {R.}~\bibnamefont {Wiesendanger}},\ }\href@noop {} {\bibfield  {journal} {\bibinfo  {journal} {New Journal of Physics}\ }\textbf {\bibinfo {volume} {18}},\ \bibinfo {pages} {045021} (\bibinfo {year} {2016})}\BibitemShut {NoStop}%
\bibitem [{\citenamefont {B{\"u}ttner}\ \emph {et~al.}(2018)\citenamefont {B{\"u}ttner}, \citenamefont {Lemesh},\ and\ \citenamefont {Beach}}]{buttner2018theory}%
  \BibitemOpen
  \bibfield  {author} {\bibinfo {author} {\bibfnamefont {F.}~\bibnamefont {B{\"u}ttner}}, \bibinfo {author} {\bibfnamefont {I.}~\bibnamefont {Lemesh}}, \ and\ \bibinfo {author} {\bibfnamefont {G.~S.}\ \bibnamefont {Beach}},\ }\href@noop {} {\bibfield  {journal} {\bibinfo  {journal} {Scientific reports}\ }\textbf {\bibinfo {volume} {8}},\ \bibinfo {pages} {4464} (\bibinfo {year} {2018})}\BibitemShut {NoStop}%
\bibitem [{\citenamefont {Jiang}\ \emph {et~al.}(2017)\citenamefont {Jiang}, \citenamefont {Zhang}, \citenamefont {Yu}, \citenamefont {Zhang}, \citenamefont {Wang}, \citenamefont {Benjamin~Jungfleisch}, \citenamefont {Pearson}, \citenamefont {Cheng}, \citenamefont {Heinonen}, \citenamefont {Wang} \emph {et~al.}}]{jiang2017direct}%
  \BibitemOpen
  \bibfield  {author} {\bibinfo {author} {\bibfnamefont {W.}~\bibnamefont {Jiang}}, \bibinfo {author} {\bibfnamefont {X.}~\bibnamefont {Zhang}}, \bibinfo {author} {\bibfnamefont {G.}~\bibnamefont {Yu}}, \bibinfo {author} {\bibfnamefont {W.}~\bibnamefont {Zhang}}, \bibinfo {author} {\bibfnamefont {X.}~\bibnamefont {Wang}}, \bibinfo {author} {\bibfnamefont {M.}~\bibnamefont {Benjamin~Jungfleisch}}, \bibinfo {author} {\bibfnamefont {J.~E.}\ \bibnamefont {Pearson}}, \bibinfo {author} {\bibfnamefont {X.}~\bibnamefont {Cheng}}, \bibinfo {author} {\bibfnamefont {O.}~\bibnamefont {Heinonen}}, \bibinfo {author} {\bibfnamefont {K.~L.}\ \bibnamefont {Wang}},  \emph {et~al.},\ }\href@noop {} {\bibfield  {journal} {\bibinfo  {journal} {Nature Physics}\ }\textbf {\bibinfo {volume} {13}},\ \bibinfo {pages} {162} (\bibinfo {year} {2017})}\BibitemShut {NoStop}%
\bibitem [{\citenamefont {Litzius}\ \emph {et~al.}(2017)\citenamefont {Litzius}, \citenamefont {Lemesh}, \citenamefont {Kr{\"u}ger}, \citenamefont {Bassirian}, \citenamefont {Caretta}, \citenamefont {Richter}, \citenamefont {B{\"u}ttner}, \citenamefont {Sato}, \citenamefont {Tretiakov}, \citenamefont {F{\"o}rster} \emph {et~al.}}]{litzius2017skyrmion}%
  \BibitemOpen
  \bibfield  {author} {\bibinfo {author} {\bibfnamefont {K.}~\bibnamefont {Litzius}}, \bibinfo {author} {\bibfnamefont {I.}~\bibnamefont {Lemesh}}, \bibinfo {author} {\bibfnamefont {B.}~\bibnamefont {Kr{\"u}ger}}, \bibinfo {author} {\bibfnamefont {P.}~\bibnamefont {Bassirian}}, \bibinfo {author} {\bibfnamefont {L.}~\bibnamefont {Caretta}}, \bibinfo {author} {\bibfnamefont {K.}~\bibnamefont {Richter}}, \bibinfo {author} {\bibfnamefont {F.}~\bibnamefont {B{\"u}ttner}}, \bibinfo {author} {\bibfnamefont {K.}~\bibnamefont {Sato}}, \bibinfo {author} {\bibfnamefont {O.~A.}\ \bibnamefont {Tretiakov}}, \bibinfo {author} {\bibfnamefont {J.}~\bibnamefont {F{\"o}rster}},  \emph {et~al.},\ }\href@noop {} {\bibfield  {journal} {\bibinfo  {journal} {Nature Physics}\ }\textbf {\bibinfo {volume} {13}},\ \bibinfo {pages} {170} (\bibinfo {year} {2017})}\BibitemShut {NoStop}%
\bibitem [{\citenamefont {Leamy}\ and\ \citenamefont {Dirks}(1979)}]{leamy1979microstructure}%
  \BibitemOpen
  \bibfield  {author} {\bibinfo {author} {\bibfnamefont {H.}~\bibnamefont {Leamy}}\ and\ \bibinfo {author} {\bibfnamefont {A.}~\bibnamefont {Dirks}},\ }\href@noop {} {\bibfield  {journal} {\bibinfo  {journal} {Journal of Applied Physics}\ }\textbf {\bibinfo {volume} {50}},\ \bibinfo {pages} {2871} (\bibinfo {year} {1979})}\BibitemShut {NoStop}%
\bibitem [{\citenamefont {Harris}\ \emph {et~al.}(1992)\citenamefont {Harris}, \citenamefont {Aylesworth}, \citenamefont {Das}, \citenamefont {Elam},\ and\ \citenamefont {Koon}}]{harris1992structural}%
  \BibitemOpen
  \bibfield  {author} {\bibinfo {author} {\bibfnamefont {V.}~\bibnamefont {Harris}}, \bibinfo {author} {\bibfnamefont {K.}~\bibnamefont {Aylesworth}}, \bibinfo {author} {\bibfnamefont {B.}~\bibnamefont {Das}}, \bibinfo {author} {\bibfnamefont {W.}~\bibnamefont {Elam}}, \ and\ \bibinfo {author} {\bibfnamefont {N.}~\bibnamefont {Koon}},\ }\href@noop {} {\bibfield  {journal} {\bibinfo  {journal} {Physical review letters}\ }\textbf {\bibinfo {volume} {69}},\ \bibinfo {pages} {1939} (\bibinfo {year} {1992})}\BibitemShut {NoStop}%
\bibitem [{\citenamefont {P.}\ and\ \citenamefont {Mohanty}(2023{\natexlab{a}})}]{P2023171158}%
  \BibitemOpen
  \bibfield  {author} {\bibinfo {author} {\bibfnamefont {S.~P.}\ \bibnamefont {P.}}\ and\ \bibinfo {author} {\bibfnamefont {J.~R.}\ \bibnamefont {Mohanty}},\ }\href@noop {} {\bibfield  {journal} {\bibinfo  {journal} {Journal of Magnetism and Magnetic Materials}\ }\textbf {\bibinfo {volume} {586}},\ \bibinfo {pages} {171158} (\bibinfo {year} {2023}{\natexlab{a}})}\BibitemShut {NoStop}%
\bibitem [{\citenamefont {Hansen}\ \emph {et~al.}(1989)\citenamefont {Hansen}, \citenamefont {Clausen}, \citenamefont {Much}, \citenamefont {Rosenkranz},\ and\ \citenamefont {Witter}}]{hansen1989magnetic}%
  \BibitemOpen
  \bibfield  {author} {\bibinfo {author} {\bibfnamefont {P.}~\bibnamefont {Hansen}}, \bibinfo {author} {\bibfnamefont {C.}~\bibnamefont {Clausen}}, \bibinfo {author} {\bibfnamefont {G.}~\bibnamefont {Much}}, \bibinfo {author} {\bibfnamefont {M.}~\bibnamefont {Rosenkranz}}, \ and\ \bibinfo {author} {\bibfnamefont {K.}~\bibnamefont {Witter}},\ }\href@noop {} {\bibfield  {journal} {\bibinfo  {journal} {Journal of applied physics}\ }\textbf {\bibinfo {volume} {66}},\ \bibinfo {pages} {756} (\bibinfo {year} {1989})}\BibitemShut {NoStop}%
\bibitem [{\citenamefont {Brand{\~a}o}\ \emph {et~al.}(2019)\citenamefont {Brand{\~a}o}, \citenamefont {Dugato}, \citenamefont {Puydinger~dos Santos},\ and\ \citenamefont {Cezar}}]{brandao2019evolution}%
  \BibitemOpen
  \bibfield  {author} {\bibinfo {author} {\bibfnamefont {J.}~\bibnamefont {Brand{\~a}o}}, \bibinfo {author} {\bibfnamefont {D.}~\bibnamefont {Dugato}}, \bibinfo {author} {\bibfnamefont {M.}~\bibnamefont {Puydinger~dos Santos}}, \ and\ \bibinfo {author} {\bibfnamefont {J.}~\bibnamefont {Cezar}},\ }\href@noop {} {\bibfield  {journal} {\bibinfo  {journal} {ACS Applied Nano Materials}\ }\textbf {\bibinfo {volume} {2}},\ \bibinfo {pages} {7532} (\bibinfo {year} {2019})}\BibitemShut {NoStop}%
\bibitem [{\citenamefont {Woo}\ \emph {et~al.}(2018)\citenamefont {Woo}, \citenamefont {Song}, \citenamefont {Zhang}, \citenamefont {Zhou}, \citenamefont {Ezawa}, \citenamefont {Liu}, \citenamefont {Finizio}, \citenamefont {Raabe}, \citenamefont {Lee}, \citenamefont {Kim} \emph {et~al.}}]{woo2018current}%
  \BibitemOpen
  \bibfield  {author} {\bibinfo {author} {\bibfnamefont {S.}~\bibnamefont {Woo}}, \bibinfo {author} {\bibfnamefont {K.~M.}\ \bibnamefont {Song}}, \bibinfo {author} {\bibfnamefont {X.}~\bibnamefont {Zhang}}, \bibinfo {author} {\bibfnamefont {Y.}~\bibnamefont {Zhou}}, \bibinfo {author} {\bibfnamefont {M.}~\bibnamefont {Ezawa}}, \bibinfo {author} {\bibfnamefont {X.}~\bibnamefont {Liu}}, \bibinfo {author} {\bibfnamefont {S.}~\bibnamefont {Finizio}}, \bibinfo {author} {\bibfnamefont {J.}~\bibnamefont {Raabe}}, \bibinfo {author} {\bibfnamefont {N.~J.}\ \bibnamefont {Lee}}, \bibinfo {author} {\bibfnamefont {S.-I.}\ \bibnamefont {Kim}},  \emph {et~al.},\ }\href@noop {} {\bibfield  {journal} {\bibinfo  {journal} {Nature communications}\ }\textbf {\bibinfo {volume} {9}},\ \bibinfo {pages} {959} (\bibinfo {year} {2018})}\BibitemShut {NoStop}%
\bibitem [{\citenamefont {Hirata}\ \emph {et~al.}(2019)\citenamefont {Hirata}, \citenamefont {Kim}, \citenamefont {Kim}, \citenamefont {Lee}, \citenamefont {Oh}, \citenamefont {Kim}, \citenamefont {Nishimura}, \citenamefont {Okuno}, \citenamefont {Futakawa}, \citenamefont {Yoshikawa} \emph {et~al.}}]{hirata2019vanishing}%
  \BibitemOpen
  \bibfield  {author} {\bibinfo {author} {\bibfnamefont {Y.}~\bibnamefont {Hirata}}, \bibinfo {author} {\bibfnamefont {D.-H.}\ \bibnamefont {Kim}}, \bibinfo {author} {\bibfnamefont {S.~K.}\ \bibnamefont {Kim}}, \bibinfo {author} {\bibfnamefont {D.-K.}\ \bibnamefont {Lee}}, \bibinfo {author} {\bibfnamefont {S.-H.}\ \bibnamefont {Oh}}, \bibinfo {author} {\bibfnamefont {D.-Y.}\ \bibnamefont {Kim}}, \bibinfo {author} {\bibfnamefont {T.}~\bibnamefont {Nishimura}}, \bibinfo {author} {\bibfnamefont {T.}~\bibnamefont {Okuno}}, \bibinfo {author} {\bibfnamefont {Y.}~\bibnamefont {Futakawa}}, \bibinfo {author} {\bibfnamefont {H.}~\bibnamefont {Yoshikawa}},  \emph {et~al.},\ }\href@noop {} {\bibfield  {journal} {\bibinfo  {journal} {Nature nanotechnology}\ }\textbf {\bibinfo {volume} {14}},\ \bibinfo {pages} {232} (\bibinfo {year} {2019})}\BibitemShut {NoStop}%
\bibitem [{\citenamefont {Iwasaki}\ \emph {et~al.}(2013{\natexlab{b}})\citenamefont {Iwasaki}, \citenamefont {Mochizuki},\ and\ \citenamefont {Nagaosa}}]{iwasaki2013universal}%
  \BibitemOpen
  \bibfield  {author} {\bibinfo {author} {\bibfnamefont {J.}~\bibnamefont {Iwasaki}}, \bibinfo {author} {\bibfnamefont {M.}~\bibnamefont {Mochizuki}}, \ and\ \bibinfo {author} {\bibfnamefont {N.}~\bibnamefont {Nagaosa}},\ }\href@noop {} {\bibfield  {journal} {\bibinfo  {journal} {Nature communications}\ }\textbf {\bibinfo {volume} {4}},\ \bibinfo {pages} {1463} (\bibinfo {year} {2013}{\natexlab{b}})}\BibitemShut {NoStop}%
\bibitem [{\citenamefont {Zhang}\ \emph {et~al.}(2018)\citenamefont {Zhang}, \citenamefont {Wang}, \citenamefont {Burn}, \citenamefont {Peng}, \citenamefont {Berger}, \citenamefont {Bauer}, \citenamefont {Pfleiderer}, \citenamefont {Van Der~Laan},\ and\ \citenamefont {Hesjedal}}]{zhang2018manipulation}%
  \BibitemOpen
  \bibfield  {author} {\bibinfo {author} {\bibfnamefont {S.}~\bibnamefont {Zhang}}, \bibinfo {author} {\bibfnamefont {W.}~\bibnamefont {Wang}}, \bibinfo {author} {\bibfnamefont {D.}~\bibnamefont {Burn}}, \bibinfo {author} {\bibfnamefont {H.}~\bibnamefont {Peng}}, \bibinfo {author} {\bibfnamefont {H.}~\bibnamefont {Berger}}, \bibinfo {author} {\bibfnamefont {A.}~\bibnamefont {Bauer}}, \bibinfo {author} {\bibfnamefont {C.}~\bibnamefont {Pfleiderer}}, \bibinfo {author} {\bibfnamefont {G.}~\bibnamefont {Van Der~Laan}}, \ and\ \bibinfo {author} {\bibfnamefont {T.}~\bibnamefont {Hesjedal}},\ }\href@noop {} {\bibfield  {journal} {\bibinfo  {journal} {Nature communications}\ }\textbf {\bibinfo {volume} {9}},\ \bibinfo {pages} {2115} (\bibinfo {year} {2018})}\BibitemShut {NoStop}%
\bibitem [{\citenamefont {Tchoe}\ and\ \citenamefont {Han}(2012)}]{tchoe2012skyrmion}%
  \BibitemOpen
  \bibfield  {author} {\bibinfo {author} {\bibfnamefont {Y.}~\bibnamefont {Tchoe}}\ and\ \bibinfo {author} {\bibfnamefont {J.~H.}\ \bibnamefont {Han}},\ }\href@noop {} {\bibfield  {journal} {\bibinfo  {journal} {Physical Review B}\ }\textbf {\bibinfo {volume} {85}},\ \bibinfo {pages} {174416} (\bibinfo {year} {2012})}\BibitemShut {NoStop}%
\bibitem [{\citenamefont {Romming}\ \emph {et~al.}(2013)\citenamefont {Romming}, \citenamefont {Hanneken}, \citenamefont {Menzel}, \citenamefont {Bickel}, \citenamefont {Wolter}, \citenamefont {von Bergmann}, \citenamefont {Kubetzka},\ and\ \citenamefont {Wiesendanger}}]{romming2013writing}%
  \BibitemOpen
  \bibfield  {author} {\bibinfo {author} {\bibfnamefont {N.}~\bibnamefont {Romming}}, \bibinfo {author} {\bibfnamefont {C.}~\bibnamefont {Hanneken}}, \bibinfo {author} {\bibfnamefont {M.}~\bibnamefont {Menzel}}, \bibinfo {author} {\bibfnamefont {J.~E.}\ \bibnamefont {Bickel}}, \bibinfo {author} {\bibfnamefont {B.}~\bibnamefont {Wolter}}, \bibinfo {author} {\bibfnamefont {K.}~\bibnamefont {von Bergmann}}, \bibinfo {author} {\bibfnamefont {A.}~\bibnamefont {Kubetzka}}, \ and\ \bibinfo {author} {\bibfnamefont {R.}~\bibnamefont {Wiesendanger}},\ }\href@noop {} {\bibfield  {journal} {\bibinfo  {journal} {Science}\ }\textbf {\bibinfo {volume} {341}},\ \bibinfo {pages} {636} (\bibinfo {year} {2013})}\BibitemShut {NoStop}%
\bibitem [{\citenamefont {Lemesh}\ \emph {et~al.}(2018)\citenamefont {Lemesh}, \citenamefont {Litzius}, \citenamefont {B{\"o}ttcher}, \citenamefont {Bassirian}, \citenamefont {Kerber}, \citenamefont {Heinze}, \citenamefont {Z{\'a}zvorka}, \citenamefont {B{\"u}ttner}, \citenamefont {Caretta}, \citenamefont {Mann} \emph {et~al.}}]{lemesh2018current}%
  \BibitemOpen
  \bibfield  {author} {\bibinfo {author} {\bibfnamefont {I.}~\bibnamefont {Lemesh}}, \bibinfo {author} {\bibfnamefont {K.}~\bibnamefont {Litzius}}, \bibinfo {author} {\bibfnamefont {M.}~\bibnamefont {B{\"o}ttcher}}, \bibinfo {author} {\bibfnamefont {P.}~\bibnamefont {Bassirian}}, \bibinfo {author} {\bibfnamefont {N.}~\bibnamefont {Kerber}}, \bibinfo {author} {\bibfnamefont {D.}~\bibnamefont {Heinze}}, \bibinfo {author} {\bibfnamefont {J.}~\bibnamefont {Z{\'a}zvorka}}, \bibinfo {author} {\bibfnamefont {F.}~\bibnamefont {B{\"u}ttner}}, \bibinfo {author} {\bibfnamefont {L.}~\bibnamefont {Caretta}}, \bibinfo {author} {\bibfnamefont {M.}~\bibnamefont {Mann}},  \emph {et~al.},\ }\href@noop {} {\bibfield  {journal} {\bibinfo  {journal} {Advanced materials}\ }\textbf {\bibinfo {volume} {30}},\ \bibinfo {pages} {1805461} (\bibinfo {year} {2018})}\BibitemShut {NoStop}%
\bibitem [{\citenamefont {Hsu}\ \emph {et~al.}(2017)\citenamefont {Hsu}, \citenamefont {Kubetzka}, \citenamefont {Finco}, \citenamefont {Romming}, \citenamefont {Von~Bergmann},\ and\ \citenamefont {Wiesendanger}}]{hsu2017electric}%
  \BibitemOpen
  \bibfield  {author} {\bibinfo {author} {\bibfnamefont {P.-J.}\ \bibnamefont {Hsu}}, \bibinfo {author} {\bibfnamefont {A.}~\bibnamefont {Kubetzka}}, \bibinfo {author} {\bibfnamefont {A.}~\bibnamefont {Finco}}, \bibinfo {author} {\bibfnamefont {N.}~\bibnamefont {Romming}}, \bibinfo {author} {\bibfnamefont {K.}~\bibnamefont {Von~Bergmann}}, \ and\ \bibinfo {author} {\bibfnamefont {R.}~\bibnamefont {Wiesendanger}},\ }\href@noop {} {\bibfield  {journal} {\bibinfo  {journal} {Nature nanotechnology}\ }\textbf {\bibinfo {volume} {12}},\ \bibinfo {pages} {123} (\bibinfo {year} {2017})}\BibitemShut {NoStop}%
\bibitem [{\citenamefont {Xia}\ \emph {et~al.}(2018)\citenamefont {Xia}, \citenamefont {Song}, \citenamefont {Jin}, \citenamefont {Wang}, \citenamefont {Wang},\ and\ \citenamefont {Liu}}]{xia2018skyrmion}%
  \BibitemOpen
  \bibfield  {author} {\bibinfo {author} {\bibfnamefont {H.}~\bibnamefont {Xia}}, \bibinfo {author} {\bibfnamefont {C.}~\bibnamefont {Song}}, \bibinfo {author} {\bibfnamefont {C.}~\bibnamefont {Jin}}, \bibinfo {author} {\bibfnamefont {J.}~\bibnamefont {Wang}}, \bibinfo {author} {\bibfnamefont {J.}~\bibnamefont {Wang}}, \ and\ \bibinfo {author} {\bibfnamefont {Q.}~\bibnamefont {Liu}},\ }\href@noop {} {\bibfield  {journal} {\bibinfo  {journal} {Journal of Magnetism and Magnetic Materials}\ }\textbf {\bibinfo {volume} {458}},\ \bibinfo {pages} {57} (\bibinfo {year} {2018})}\BibitemShut {NoStop}%
\bibitem [{\citenamefont {Gorshkov}\ \emph {et~al.}(2022)\citenamefont {Gorshkov}, \citenamefont {Gorev}, \citenamefont {Sapozhnikov},\ and\ \citenamefont {Udalov}}]{gorshkov2022dmi}%
  \BibitemOpen
  \bibfield  {author} {\bibinfo {author} {\bibfnamefont {I.~O.}\ \bibnamefont {Gorshkov}}, \bibinfo {author} {\bibfnamefont {R.~V.}\ \bibnamefont {Gorev}}, \bibinfo {author} {\bibfnamefont {M.~V.}\ \bibnamefont {Sapozhnikov}}, \ and\ \bibinfo {author} {\bibfnamefont {O.~G.}\ \bibnamefont {Udalov}},\ }\href@noop {} {\bibfield  {journal} {\bibinfo  {journal} {ACS Applied Electronic Materials}\ }\textbf {\bibinfo {volume} {4}},\ \bibinfo {pages} {3205} (\bibinfo {year} {2022})}\BibitemShut {NoStop}%
\bibitem [{\citenamefont {Beaurepaire}\ \emph {et~al.}(1996)\citenamefont {Beaurepaire}, \citenamefont {Merle}, \citenamefont {Daunois},\ and\ \citenamefont {Bigot}}]{beaurepaire1996ultrafast}%
  \BibitemOpen
  \bibfield  {author} {\bibinfo {author} {\bibfnamefont {E.}~\bibnamefont {Beaurepaire}}, \bibinfo {author} {\bibfnamefont {J.-C.}\ \bibnamefont {Merle}}, \bibinfo {author} {\bibfnamefont {A.}~\bibnamefont {Daunois}}, \ and\ \bibinfo {author} {\bibfnamefont {J.-Y.}\ \bibnamefont {Bigot}},\ }\href@noop {} {\bibfield  {journal} {\bibinfo  {journal} {Physical review letters}\ }\textbf {\bibinfo {volume} {76}},\ \bibinfo {pages} {4250} (\bibinfo {year} {1996})}\BibitemShut {NoStop}%
\bibitem [{\citenamefont {Stanciu}\ \emph {et~al.}(2007)\citenamefont {Stanciu}, \citenamefont {Hansteen}, \citenamefont {Kimel}, \citenamefont {Kirilyuk}, \citenamefont {Tsukamoto}, \citenamefont {Itoh},\ and\ \citenamefont {Rasing}}]{stanciu2007all}%
  \BibitemOpen
  \bibfield  {author} {\bibinfo {author} {\bibfnamefont {C.~D.}\ \bibnamefont {Stanciu}}, \bibinfo {author} {\bibfnamefont {F.}~\bibnamefont {Hansteen}}, \bibinfo {author} {\bibfnamefont {A.~V.}\ \bibnamefont {Kimel}}, \bibinfo {author} {\bibfnamefont {A.}~\bibnamefont {Kirilyuk}}, \bibinfo {author} {\bibfnamefont {A.}~\bibnamefont {Tsukamoto}}, \bibinfo {author} {\bibfnamefont {A.}~\bibnamefont {Itoh}}, \ and\ \bibinfo {author} {\bibfnamefont {T.}~\bibnamefont {Rasing}},\ }\href@noop {} {\bibfield  {journal} {\bibinfo  {journal} {Physical review letters}\ }\textbf {\bibinfo {volume} {99}},\ \bibinfo {pages} {047601} (\bibinfo {year} {2007})}\BibitemShut {NoStop}%
\bibitem [{\citenamefont {Radu}\ \emph {et~al.}(2011)\citenamefont {Radu}, \citenamefont {Vahaplar}, \citenamefont {Stamm}, \citenamefont {Kachel}, \citenamefont {Pontius}, \citenamefont {D{\"u}rr}, \citenamefont {Ostler}, \citenamefont {Barker}, \citenamefont {Evans}, \citenamefont {Chantrell} \emph {et~al.}}]{radu2011transient}%
  \BibitemOpen
  \bibfield  {author} {\bibinfo {author} {\bibfnamefont {I.}~\bibnamefont {Radu}}, \bibinfo {author} {\bibfnamefont {K.}~\bibnamefont {Vahaplar}}, \bibinfo {author} {\bibfnamefont {C.}~\bibnamefont {Stamm}}, \bibinfo {author} {\bibfnamefont {T.}~\bibnamefont {Kachel}}, \bibinfo {author} {\bibfnamefont {N.}~\bibnamefont {Pontius}}, \bibinfo {author} {\bibfnamefont {H.}~\bibnamefont {D{\"u}rr}}, \bibinfo {author} {\bibfnamefont {T.}~\bibnamefont {Ostler}}, \bibinfo {author} {\bibfnamefont {J.}~\bibnamefont {Barker}}, \bibinfo {author} {\bibfnamefont {R.}~\bibnamefont {Evans}}, \bibinfo {author} {\bibfnamefont {R.}~\bibnamefont {Chantrell}},  \emph {et~al.},\ }\href@noop {} {\bibfield  {journal} {\bibinfo  {journal} {Nature}\ }\textbf {\bibinfo {volume} {472}},\ \bibinfo {pages} {205} (\bibinfo {year} {2011})}\BibitemShut {NoStop}%
\bibitem [{\citenamefont {Ostler}\ \emph {et~al.}(2012)\citenamefont {Ostler}, \citenamefont {Barker}, \citenamefont {Evans}, \citenamefont {Chantrell}, \citenamefont {Atxitia}, \citenamefont {Chubykalo-Fesenko}, \citenamefont {El~Moussaoui}, \citenamefont {Le~Guyader}, \citenamefont {Mengotti}, \citenamefont {Heyderman} \emph {et~al.}}]{ostler2012ultrafast}%
  \BibitemOpen
  \bibfield  {author} {\bibinfo {author} {\bibfnamefont {T.}~\bibnamefont {Ostler}}, \bibinfo {author} {\bibfnamefont {J.}~\bibnamefont {Barker}}, \bibinfo {author} {\bibfnamefont {R.}~\bibnamefont {Evans}}, \bibinfo {author} {\bibfnamefont {R.}~\bibnamefont {Chantrell}}, \bibinfo {author} {\bibfnamefont {U.}~\bibnamefont {Atxitia}}, \bibinfo {author} {\bibfnamefont {O.}~\bibnamefont {Chubykalo-Fesenko}}, \bibinfo {author} {\bibfnamefont {S.}~\bibnamefont {El~Moussaoui}}, \bibinfo {author} {\bibfnamefont {L.}~\bibnamefont {Le~Guyader}}, \bibinfo {author} {\bibfnamefont {E.}~\bibnamefont {Mengotti}}, \bibinfo {author} {\bibfnamefont {L.}~\bibnamefont {Heyderman}},  \emph {et~al.},\ }\href@noop {} {\bibfield  {journal} {\bibinfo  {journal} {Nature communications}\ }\textbf {\bibinfo {volume} {3}},\ \bibinfo {pages} {1} (\bibinfo {year} {2012})}\BibitemShut {NoStop}%
\bibitem [{\citenamefont {Mangin}\ \emph {et~al.}(2014)\citenamefont {Mangin}, \citenamefont {Gottwald}, \citenamefont {Lambert}, \citenamefont {Steil}, \citenamefont {Uhl{\'\i}{\v{r}}}, \citenamefont {Pang}, \citenamefont {Hehn}, \citenamefont {Alebrand}, \citenamefont {Cinchetti}, \citenamefont {Malinowski} \emph {et~al.}}]{mangin2014engineered}%
  \BibitemOpen
  \bibfield  {author} {\bibinfo {author} {\bibfnamefont {S.}~\bibnamefont {Mangin}}, \bibinfo {author} {\bibfnamefont {M.}~\bibnamefont {Gottwald}}, \bibinfo {author} {\bibfnamefont {C.}~\bibnamefont {Lambert}}, \bibinfo {author} {\bibfnamefont {D.}~\bibnamefont {Steil}}, \bibinfo {author} {\bibfnamefont {V.}~\bibnamefont {Uhl{\'\i}{\v{r}}}}, \bibinfo {author} {\bibfnamefont {L.}~\bibnamefont {Pang}}, \bibinfo {author} {\bibfnamefont {M.}~\bibnamefont {Hehn}}, \bibinfo {author} {\bibfnamefont {S.}~\bibnamefont {Alebrand}}, \bibinfo {author} {\bibfnamefont {M.}~\bibnamefont {Cinchetti}}, \bibinfo {author} {\bibfnamefont {G.}~\bibnamefont {Malinowski}},  \emph {et~al.},\ }\href@noop {} {\bibfield  {journal} {\bibinfo  {journal} {Nature materials}\ }\textbf {\bibinfo {volume} {13}},\ \bibinfo {pages} {286} (\bibinfo {year} {2014})}\BibitemShut {NoStop}%
\bibitem [{\citenamefont {Jakobs}\ \emph {et~al.}(2021)\citenamefont {Jakobs}, \citenamefont {Ostler}, \citenamefont {Lambert}, \citenamefont {Yang}, \citenamefont {Salahuddin}, \citenamefont {Wilson}, \citenamefont {Gorchon}, \citenamefont {Bokor},\ and\ \citenamefont {Atxitia}}]{jakobs2021unifying}%
  \BibitemOpen
  \bibfield  {author} {\bibinfo {author} {\bibfnamefont {F.}~\bibnamefont {Jakobs}}, \bibinfo {author} {\bibfnamefont {T.}~\bibnamefont {Ostler}}, \bibinfo {author} {\bibfnamefont {C.-H.}\ \bibnamefont {Lambert}}, \bibinfo {author} {\bibfnamefont {Y.}~\bibnamefont {Yang}}, \bibinfo {author} {\bibfnamefont {S.}~\bibnamefont {Salahuddin}}, \bibinfo {author} {\bibfnamefont {R.~B.}\ \bibnamefont {Wilson}}, \bibinfo {author} {\bibfnamefont {J.}~\bibnamefont {Gorchon}}, \bibinfo {author} {\bibfnamefont {J.}~\bibnamefont {Bokor}}, \ and\ \bibinfo {author} {\bibfnamefont {U.}~\bibnamefont {Atxitia}},\ }\href@noop {} {\bibfield  {journal} {\bibinfo  {journal} {Physical Review B}\ }\textbf {\bibinfo {volume} {103}},\ \bibinfo {pages} {104422} (\bibinfo {year} {2021})}\BibitemShut {NoStop}%
\bibitem [{\citenamefont {Gweha~Nyoma}\ \emph {et~al.}(2024)\citenamefont {Gweha~Nyoma}, \citenamefont {Hehn}, \citenamefont {Malinowski}, \citenamefont {Ghanbaja}, \citenamefont {Hohlfeld}, \citenamefont {Gorchon}, \citenamefont {Mangin},\ and\ \citenamefont {Montaigne}}]{gweha2024gd}%
  \BibitemOpen
  \bibfield  {author} {\bibinfo {author} {\bibfnamefont {D.~P.}\ \bibnamefont {Gweha~Nyoma}}, \bibinfo {author} {\bibfnamefont {M.}~\bibnamefont {Hehn}}, \bibinfo {author} {\bibfnamefont {G.}~\bibnamefont {Malinowski}}, \bibinfo {author} {\bibfnamefont {J.}~\bibnamefont {Ghanbaja}}, \bibinfo {author} {\bibfnamefont {J.}~\bibnamefont {Hohlfeld}}, \bibinfo {author} {\bibfnamefont {J.}~\bibnamefont {Gorchon}}, \bibinfo {author} {\bibfnamefont {S.}~\bibnamefont {Mangin}}, \ and\ \bibinfo {author} {\bibfnamefont {F.}~\bibnamefont {Montaigne}},\ }\href@noop {} {\bibfield  {journal} {\bibinfo  {journal} {ACS Applied Electronic Materials}\ } (\bibinfo {year} {2024})}\BibitemShut {NoStop}%
\bibitem [{\citenamefont {Koshibae}\ and\ \citenamefont {Nagaosa}(2014)}]{koshibae2014creation}%
  \BibitemOpen
  \bibfield  {author} {\bibinfo {author} {\bibfnamefont {W.}~\bibnamefont {Koshibae}}\ and\ \bibinfo {author} {\bibfnamefont {N.}~\bibnamefont {Nagaosa}},\ }\href@noop {} {\bibfield  {journal} {\bibinfo  {journal} {Nature communications}\ }\textbf {\bibinfo {volume} {5}},\ \bibinfo {pages} {5148} (\bibinfo {year} {2014})}\BibitemShut {NoStop}%
\bibitem [{\citenamefont {Je}\ \emph {et~al.}(2018)\citenamefont {Je}, \citenamefont {Vallobra}, \citenamefont {Srivastava}, \citenamefont {Rojas-S{\'a}nchez}, \citenamefont {Pham}, \citenamefont {Hehn}, \citenamefont {Malinowski}, \citenamefont {Baraduc}, \citenamefont {Auffret}, \citenamefont {Gaudin} \emph {et~al.}}]{je2018creation}%
  \BibitemOpen
  \bibfield  {author} {\bibinfo {author} {\bibfnamefont {S.-G.}\ \bibnamefont {Je}}, \bibinfo {author} {\bibfnamefont {P.}~\bibnamefont {Vallobra}}, \bibinfo {author} {\bibfnamefont {T.}~\bibnamefont {Srivastava}}, \bibinfo {author} {\bibfnamefont {J.-C.}\ \bibnamefont {Rojas-S{\'a}nchez}}, \bibinfo {author} {\bibfnamefont {T.~H.}\ \bibnamefont {Pham}}, \bibinfo {author} {\bibfnamefont {M.}~\bibnamefont {Hehn}}, \bibinfo {author} {\bibfnamefont {G.}~\bibnamefont {Malinowski}}, \bibinfo {author} {\bibfnamefont {C.}~\bibnamefont {Baraduc}}, \bibinfo {author} {\bibfnamefont {S.}~\bibnamefont {Auffret}}, \bibinfo {author} {\bibfnamefont {G.}~\bibnamefont {Gaudin}},  \emph {et~al.},\ }\href@noop {} {\bibfield  {journal} {\bibinfo  {journal} {Nano letters}\ }\textbf {\bibinfo {volume} {18}},\ \bibinfo {pages} {7362} (\bibinfo {year} {2018})}\BibitemShut {NoStop}%
\bibitem [{\citenamefont {Finazzi}\ \emph {et~al.}(2013)\citenamefont {Finazzi}, \citenamefont {Savoini}, \citenamefont {Khorsand}, \citenamefont {Tsukamoto}, \citenamefont {Itoh}, \citenamefont {Duo}, \citenamefont {Kirilyuk}, \citenamefont {Rasing},\ and\ \citenamefont {Ezawa}}]{finazzi2013laser}%
  \BibitemOpen
  \bibfield  {author} {\bibinfo {author} {\bibfnamefont {M.}~\bibnamefont {Finazzi}}, \bibinfo {author} {\bibfnamefont {M.}~\bibnamefont {Savoini}}, \bibinfo {author} {\bibfnamefont {A.}~\bibnamefont {Khorsand}}, \bibinfo {author} {\bibfnamefont {A.}~\bibnamefont {Tsukamoto}}, \bibinfo {author} {\bibfnamefont {A.}~\bibnamefont {Itoh}}, \bibinfo {author} {\bibfnamefont {L.}~\bibnamefont {Duo}}, \bibinfo {author} {\bibfnamefont {A.}~\bibnamefont {Kirilyuk}}, \bibinfo {author} {\bibfnamefont {T.}~\bibnamefont {Rasing}}, \ and\ \bibinfo {author} {\bibfnamefont {M.}~\bibnamefont {Ezawa}},\ }\href@noop {} {\bibfield  {journal} {\bibinfo  {journal} {Physical review letters}\ }\textbf {\bibinfo {volume} {110}},\ \bibinfo {pages} {177205} (\bibinfo {year} {2013})}\BibitemShut {NoStop}%
\bibitem [{\citenamefont {Berruto}\ \emph {et~al.}(2018)\citenamefont {Berruto}, \citenamefont {Madan}, \citenamefont {Murooka}, \citenamefont {Vanacore}, \citenamefont {Pomarico}, \citenamefont {Rajeswari}, \citenamefont {Lamb}, \citenamefont {Huang}, \citenamefont {Kruchkov}, \citenamefont {Togawa} \emph {et~al.}}]{berruto2018laser}%
  \BibitemOpen
  \bibfield  {author} {\bibinfo {author} {\bibfnamefont {G.}~\bibnamefont {Berruto}}, \bibinfo {author} {\bibfnamefont {I.}~\bibnamefont {Madan}}, \bibinfo {author} {\bibfnamefont {Y.}~\bibnamefont {Murooka}}, \bibinfo {author} {\bibfnamefont {G.}~\bibnamefont {Vanacore}}, \bibinfo {author} {\bibfnamefont {E.}~\bibnamefont {Pomarico}}, \bibinfo {author} {\bibfnamefont {J.}~\bibnamefont {Rajeswari}}, \bibinfo {author} {\bibfnamefont {R.}~\bibnamefont {Lamb}}, \bibinfo {author} {\bibfnamefont {P.}~\bibnamefont {Huang}}, \bibinfo {author} {\bibfnamefont {A.}~\bibnamefont {Kruchkov}}, \bibinfo {author} {\bibfnamefont {Y.}~\bibnamefont {Togawa}},  \emph {et~al.},\ }\href@noop {} {\bibfield  {journal} {\bibinfo  {journal} {Physical review letters}\ }\textbf {\bibinfo {volume} {120}},\ \bibinfo {pages} {117201} (\bibinfo {year} {2018})}\BibitemShut {NoStop}%
\bibitem [{\citenamefont {B{\"u}ttner}\ \emph {et~al.}(2021)\citenamefont {B{\"u}ttner}, \citenamefont {Pfau}, \citenamefont {B{\"o}ttcher}, \citenamefont {Schneider}, \citenamefont {Mercurio}, \citenamefont {G{\"u}nther}, \citenamefont {Hessing}, \citenamefont {Klose}, \citenamefont {Wittmann}, \citenamefont {Gerlinger} \emph {et~al.}}]{buttner2021observation}%
  \BibitemOpen
  \bibfield  {author} {\bibinfo {author} {\bibfnamefont {F.}~\bibnamefont {B{\"u}ttner}}, \bibinfo {author} {\bibfnamefont {B.}~\bibnamefont {Pfau}}, \bibinfo {author} {\bibfnamefont {M.}~\bibnamefont {B{\"o}ttcher}}, \bibinfo {author} {\bibfnamefont {M.}~\bibnamefont {Schneider}}, \bibinfo {author} {\bibfnamefont {G.}~\bibnamefont {Mercurio}}, \bibinfo {author} {\bibfnamefont {C.~M.}\ \bibnamefont {G{\"u}nther}}, \bibinfo {author} {\bibfnamefont {P.}~\bibnamefont {Hessing}}, \bibinfo {author} {\bibfnamefont {C.}~\bibnamefont {Klose}}, \bibinfo {author} {\bibfnamefont {A.}~\bibnamefont {Wittmann}}, \bibinfo {author} {\bibfnamefont {K.}~\bibnamefont {Gerlinger}},  \emph {et~al.},\ }\href@noop {} {\bibfield  {journal} {\bibinfo  {journal} {Nature materials}\ }\textbf {\bibinfo {volume} {20}},\ \bibinfo {pages} {30} (\bibinfo {year} {2021})}\BibitemShut {NoStop}%
\bibitem [{\citenamefont {Zhang}\ \emph {et~al.}(2023)\citenamefont {Zhang}, \citenamefont {Huang}, \citenamefont {Hehn}, \citenamefont {Malinowski}, \citenamefont {Verges}, \citenamefont {Hohlfeld}, \citenamefont {Remy}, \citenamefont {Lacour}, \citenamefont {Wang}, \citenamefont {Zhao} \emph {et~al.}}]{zhang2023optical}%
  \BibitemOpen
  \bibfield  {author} {\bibinfo {author} {\bibfnamefont {W.}~\bibnamefont {Zhang}}, \bibinfo {author} {\bibfnamefont {T.~X.}\ \bibnamefont {Huang}}, \bibinfo {author} {\bibfnamefont {M.}~\bibnamefont {Hehn}}, \bibinfo {author} {\bibfnamefont {G.}~\bibnamefont {Malinowski}}, \bibinfo {author} {\bibfnamefont {M.}~\bibnamefont {Verges}}, \bibinfo {author} {\bibfnamefont {J.}~\bibnamefont {Hohlfeld}}, \bibinfo {author} {\bibfnamefont {Q.}~\bibnamefont {Remy}}, \bibinfo {author} {\bibfnamefont {D.}~\bibnamefont {Lacour}}, \bibinfo {author} {\bibfnamefont {X.~R.}\ \bibnamefont {Wang}}, \bibinfo {author} {\bibfnamefont {G.~P.}\ \bibnamefont {Zhao}},  \emph {et~al.},\ }\href@noop {} {\bibfield  {journal} {\bibinfo  {journal} {ACS Applied Materials \& Interfaces}\ }\textbf {\bibinfo {volume} {15}},\ \bibinfo {pages} {5608} (\bibinfo {year} {2023})}\BibitemShut {NoStop}%
\bibitem [{\citenamefont {Iacocca}\ \emph {et~al.}(2019)\citenamefont {Iacocca}, \citenamefont {Liu}, \citenamefont {Reid}, \citenamefont {Fu}, \citenamefont {Ruta}, \citenamefont {Granitzka}, \citenamefont {Jal}, \citenamefont {Bonetti}, \citenamefont {Gray}, \citenamefont {Graves} \emph {et~al.}}]{iacocca2019spin}%
  \BibitemOpen
  \bibfield  {author} {\bibinfo {author} {\bibfnamefont {E.}~\bibnamefont {Iacocca}}, \bibinfo {author} {\bibfnamefont {T.-M.}\ \bibnamefont {Liu}}, \bibinfo {author} {\bibfnamefont {A.~H.}\ \bibnamefont {Reid}}, \bibinfo {author} {\bibfnamefont {Z.}~\bibnamefont {Fu}}, \bibinfo {author} {\bibfnamefont {S.}~\bibnamefont {Ruta}}, \bibinfo {author} {\bibfnamefont {P.}~\bibnamefont {Granitzka}}, \bibinfo {author} {\bibfnamefont {E.}~\bibnamefont {Jal}}, \bibinfo {author} {\bibfnamefont {S.}~\bibnamefont {Bonetti}}, \bibinfo {author} {\bibfnamefont {A.}~\bibnamefont {Gray}}, \bibinfo {author} {\bibfnamefont {C.}~\bibnamefont {Graves}},  \emph {et~al.},\ }\href@noop {} {\bibfield  {journal} {\bibinfo  {journal} {Nature communications}\ }\textbf {\bibinfo {volume} {10}},\ \bibinfo {pages} {1756} (\bibinfo {year} {2019})}\BibitemShut {NoStop}%
\bibitem [{\citenamefont {Barker}\ \emph {et~al.}(2013)\citenamefont {Barker}, \citenamefont {Atxitia}, \citenamefont {Ostler}, \citenamefont {Hovorka}, \citenamefont {Chubykalo-Fesenko},\ and\ \citenamefont {Chantrell}}]{barker2013two}%
  \BibitemOpen
  \bibfield  {author} {\bibinfo {author} {\bibfnamefont {J.}~\bibnamefont {Barker}}, \bibinfo {author} {\bibfnamefont {U.}~\bibnamefont {Atxitia}}, \bibinfo {author} {\bibfnamefont {T.}~\bibnamefont {Ostler}}, \bibinfo {author} {\bibfnamefont {O.}~\bibnamefont {Hovorka}}, \bibinfo {author} {\bibfnamefont {O.}~\bibnamefont {Chubykalo-Fesenko}}, \ and\ \bibinfo {author} {\bibfnamefont {R.}~\bibnamefont {Chantrell}},\ }\href@noop {} {\bibfield  {journal} {\bibinfo  {journal} {Scientific reports}\ }\textbf {\bibinfo {volume} {3}},\ \bibinfo {pages} {3262} (\bibinfo {year} {2013})}\BibitemShut {NoStop}%
\bibitem [{\citenamefont {P.}\ and\ \citenamefont {Mohanty}(2023{\natexlab{b}})}]{P2023170701}%
  \BibitemOpen
  \bibfield  {author} {\bibinfo {author} {\bibfnamefont {S.~P.}\ \bibnamefont {P.}}\ and\ \bibinfo {author} {\bibfnamefont {J.~R.}\ \bibnamefont {Mohanty}},\ }\href@noop {} {\bibfield  {journal} {\bibinfo  {journal} {Journal of Magnetism and Magnetic Materials}\ }\textbf {\bibinfo {volume} {575}},\ \bibinfo {pages} {170701} (\bibinfo {year} {2023}{\natexlab{b}})}\BibitemShut {NoStop}%
\bibitem [{\citenamefont {Olleros-Rodr{\'\i}guez}\ \emph {et~al.}(2022)\citenamefont {Olleros-Rodr{\'\i}guez}, \citenamefont {Strungaru}, \citenamefont {Ruta}, \citenamefont {Gavriloaea}, \citenamefont {Gud{\'\i}n}, \citenamefont {Perna}, \citenamefont {Chantrell},\ and\ \citenamefont {Chubykalo-Fesenko}}]{olleros2022non}%
  \BibitemOpen
  \bibfield  {author} {\bibinfo {author} {\bibfnamefont {P.}~\bibnamefont {Olleros-Rodr{\'\i}guez}}, \bibinfo {author} {\bibfnamefont {M.}~\bibnamefont {Strungaru}}, \bibinfo {author} {\bibfnamefont {S.}~\bibnamefont {Ruta}}, \bibinfo {author} {\bibfnamefont {P.-I.}\ \bibnamefont {Gavriloaea}}, \bibinfo {author} {\bibfnamefont {A.}~\bibnamefont {Gud{\'\i}n}}, \bibinfo {author} {\bibfnamefont {P.}~\bibnamefont {Perna}}, \bibinfo {author} {\bibfnamefont {R.}~\bibnamefont {Chantrell}}, \ and\ \bibinfo {author} {\bibfnamefont {O.}~\bibnamefont {Chubykalo-Fesenko}},\ }\href@noop {} {\bibfield  {journal} {\bibinfo  {journal} {Nanoscale}\ }\textbf {\bibinfo {volume} {14}},\ \bibinfo {pages} {15701} (\bibinfo {year} {2022})}\BibitemShut {NoStop}%
\bibitem [{\citenamefont {R{\'o}zsa}\ \emph {et~al.}(2016)\citenamefont {R{\'o}zsa}, \citenamefont {Simon}, \citenamefont {Palot{\'a}s}, \citenamefont {Udvardi},\ and\ \citenamefont {Szunyogh}}]{rozsa2016complex}%
  \BibitemOpen
  \bibfield  {author} {\bibinfo {author} {\bibfnamefont {L.}~\bibnamefont {R{\'o}zsa}}, \bibinfo {author} {\bibfnamefont {E.}~\bibnamefont {Simon}}, \bibinfo {author} {\bibfnamefont {K.}~\bibnamefont {Palot{\'a}s}}, \bibinfo {author} {\bibfnamefont {L.}~\bibnamefont {Udvardi}}, \ and\ \bibinfo {author} {\bibfnamefont {L.}~\bibnamefont {Szunyogh}},\ }\href@noop {} {\bibfield  {journal} {\bibinfo  {journal} {Physical Review B}\ }\textbf {\bibinfo {volume} {93}},\ \bibinfo {pages} {024417} (\bibinfo {year} {2016})}\BibitemShut {NoStop}%
\bibitem [{\citenamefont {Evans}\ \emph {et~al.}(2014)\citenamefont {Evans}, \citenamefont {Fan}, \citenamefont {Chureemart}, \citenamefont {Ostler}, \citenamefont {Ellis},\ and\ \citenamefont {Chantrell}}]{evans2014atomistic}%
  \BibitemOpen
  \bibfield  {author} {\bibinfo {author} {\bibfnamefont {R.~F.}\ \bibnamefont {Evans}}, \bibinfo {author} {\bibfnamefont {W.~J.}\ \bibnamefont {Fan}}, \bibinfo {author} {\bibfnamefont {P.}~\bibnamefont {Chureemart}}, \bibinfo {author} {\bibfnamefont {T.~A.}\ \bibnamefont {Ostler}}, \bibinfo {author} {\bibfnamefont {M.~O.}\ \bibnamefont {Ellis}}, \ and\ \bibinfo {author} {\bibfnamefont {R.~W.}\ \bibnamefont {Chantrell}},\ }\href@noop {} {\bibfield  {journal} {\bibinfo  {journal} {Journal of Physics: Condensed Matter}\ }\textbf {\bibinfo {volume} {26}},\ \bibinfo {pages} {103202} (\bibinfo {year} {2014})}\BibitemShut {NoStop}%
\end{thebibliography}%

\end{document}